\documentclass[preprint,12pt]{elsarticle}

\usepackage{amssymb}
\usepackage{amsmath}
\usepackage{booktabs}
\usepackage{caption}
\usepackage{subcaption}
\usepackage{multirow}
\usepackage{floatrow}
\usepackage{hyperref}
\usepackage{url}
\usepackage{graphicx}
\usepackage{makecell}
\usepackage{xcolor, array, colortbl}
\usepackage{graphicx}
\usepackage[ruled, linesnumbered]{algorithm2e}
\usepackage{arydshln}
\usepackage{utfsym}

\newcommand{\lz}[1]{{\color{black}#1}}
\newcommand{\ry}[1]{{\color{black}#1}}

\definecolor{deepred}{rgb}{0.698,0.133,0.133}
\definecolor{blue}{rgb}{0,0,1}
\definecolor{highlightgreen}{HTML}{009901}
\definecolor{highlightred}{HTML}{FD6864}
\definecolor{shadecolor}{rgb}{0.92,0.92,0.92}
\definecolor{LightCyan1}{rgb}{0.88,1,1}
\journal{Artificial Intelligence}

\begin{document}

\begin{frontmatter}



% \title{P2Mark: Plug-and-play Parameter Intrinsic Watermarking for Speech Generation System Protection}
\title{P2Mark: Plug-and-play Parameter-level Watermarking for Neural Speech Generation}

\author[label1,label2]{Yong Ren}
\author[label3]{Jiangyan Yi\corref{mycorrespondingauthor}}
% \ead{yijy@mail.tsinghua.edu.cn}
\author[label1]{Tao Wang}
\author[label3,label4]{Jianhua Tao\corref{mycorrespondingauthor}}
% \ead{jhtao@tsinghua.edu.cn}
\author[label1]{Zheng Lian}
\author[label4]{Zhengqi Wen\corref{mycorrespondingauthor}}
% \ead{zqwen@tsinghua.edu.cn}
\author[label1]{Chenxing Li}
\author[label1]{Ruibo Fu}
% \author[label1,label2]{Junzuo Zhou}
% \author[label1,label2]{Hao Gu}
% \author[label5]{Xinrui Yan}
\author[label1]{Ye Bai}
\author[label1]{Xiaohui Zhang}
% \cortext[mycorrespondingauthor]{Corresponding author.}
\cortext[mycorrespondingauthor]{Corresponding authors. Email addresses: jhtao@tsinghua.edu.cn (Jianhua Tao), yijy@mail.tsinghua.edu.cn (Jiangyan Yi), zqwen@tsinghua.edu.cn (Zhengqi Wen)}

% \ead{jhtao@tsinghua.edu.cn}
% \ead{yijy@mail.tsinghua.edu.cn}
% \ead{zqwen@tsinghua.edu.cn}

\affiliation[label1]{
	organization={The State Key Laboratory of Multimodal Artificial Intelligence Systems, Institute of Automation, Chinese Academy of Sciences},
	city={Beijing},
	country={China}
}
\affiliation[label2]{
	organization={School of Artificial Intelligence, University of Chinese Academy of Sciences},
	city={Beijing},
	country={China}
}
\affiliation[label3]{
	organization={Department of Automation, Tsinghua University},
	city={Beijing},
	country={China}
}
\affiliation[label4]{
	organization={Beijing National Research Center for Information Science and Technology, Tsinghua University},
	city={Beijing},
	country={China}
}
% \affiliation[label5]{
%     organization={Beihang University},
%     city={Beijing},
% 	country={China}
% }

%% Abstract
\begin{abstract}

\ry{

Neural speech generation (NSG) has rapidly advanced as a key component of artificial intelligence-generated content, enabling the generation of high-quality, highly realistic speech for diverse applications. This development increases the risk of technique misuse and threatens social security. Audio watermarking can embed imperceptible marks into generated audio, providing a promising approach for secure NSG usage. However, current audio watermarking methods are mainly applied at the audio-level or feature-level, which are not suitable for open-sourced scenarios where source codes and model weights are released. To address this limitation, we propose a \emph{Plug-and-play Parameter-level WaterMarking (P2Mark)} method for NSG. Specifically, we embed watermarks into the released model weights, offering a reliable solution for proactively tracing and protecting model copyrights in open-source scenarios. During training, we introduce a lightweight watermark adapter into the pre-trained model, allowing watermark information to be merged into the model via this adapter. This design ensures both the flexibility to modify the watermark before model release and the security of embedding the watermark within model parameters after model release. Meanwhile, we propose a gradient orthogonal projection optimization strategy to ensure the quality of the generated audio and the accuracy of watermark preservation. Experimental results on two mainstream waveform decoders in NSG (i.e., vocoder and codec) demonstrate that P2Mark achieves comparable performance to state-of-the-art audio watermarking methods that are not applicable to open-source white-box protection scenarios, in terms of watermark extraction accuracy, watermark imperceptibility, and robustness.

}

\end{abstract}


% \begin{highlights}
% \item 
% Plug-and-play watermarking module for neural speech generation.
% \item 
% Parameter-level watermarking fusion mechanism for white-box protection.
% \item 
% Watermarking gradient orthogonal projection optimization method.
% \item 
% The effectiveness was verified on two types of speech decoders: vocoder and codec.
% \end{highlights}

%% Keywords

\end{frontmatter}


\section{Introduction}
% Recently, the latest advancements in generative models have significantly propelled the development of neural speech generation. Some of the latest speech generation methods like CosyVoice 1/2\cite{du2024cosyvoice,du2024cosyvoice2}, MaskGCT\cite{wang2024maskgct}, and Spark-TTS\cite{wang2025sparktts} are capable of producing speech that is comparable to natural human speech and have made their source codes and model weights available as open-source. The open-sourcing of these speech generation models not only provides users with a convenient way to generate high-quality personalized speech but also promotes the dissemination and advancement of speech generation technology.

% As these open-source neural speech generation models become increasingly prevalent, the risk of malicious use also increases. Firstly, some users with malicious intent may exploit powerful speech generation models to create realistic voices for fraudulent and other illegal purposes, threatening personal property security and social stability\cite{yamagishi2021asvspoof,yi2022add,yi2023add}. Secondly, it is crucial to protect the copyrights of open-source models and the intellectual property of developers. Utilizing open-source models for profit without the consent of the developers infringes on their rights. Currently, there is a lack of effective protection mechanisms for open-source speech generation models, making it difficult to trace the attribution of the speech generated by open-source models.

\lz{

Recently, the rapid advancements in generative models have significantly propelled the development of neural speech generation (NSG). Some of the excellent NSG systems \cite{du2024cosyvoice,du2024cosyvoice2,wang2024maskgct,wang2025sparktts} can generate high-quality human-like speech, and it is difficult for ordinary people to distinguish synthetic speech from real one. Although these techniques can be applied in many areas, such as achieving more realistic human-computer interaction, the malicious use of these techniques may also threaten property security and social stability. For example, some users with malicious intent may exploit NSG techniques to create realistic voices for fraudulent and other illegal purposes \cite{yamagishi2021asvspoof,yi2022add,yi2023add}.

}

\begin{figure*}[!t]
\centering
  \includegraphics[width=\linewidth]{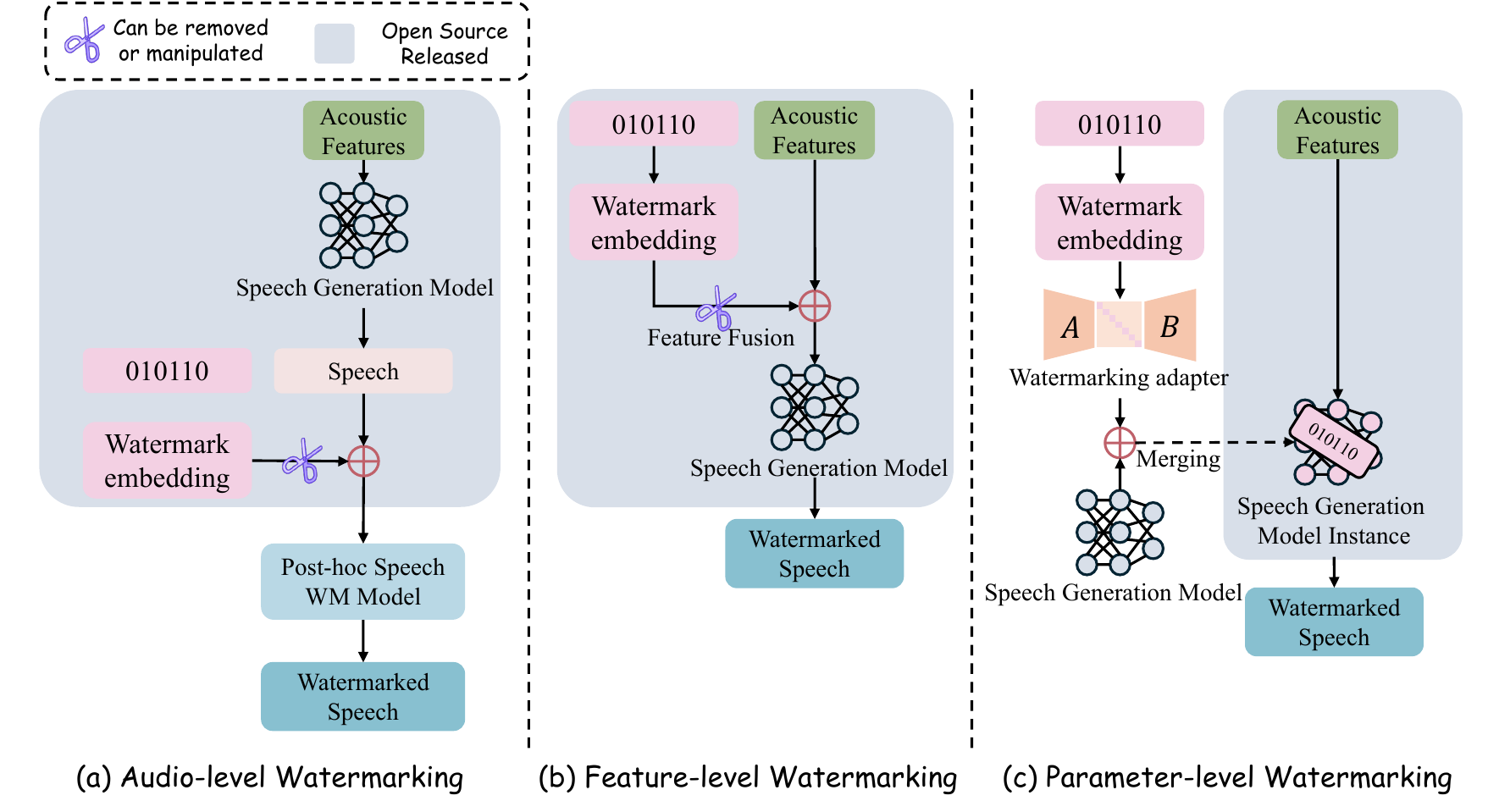} 
\caption{
% Types of audio watermarking methods. Audio watermarks can be divided into (a) Post-hoc audio watermarking methods and (b) generative model audio watermarking methods. Existing generative model audio watermarking methods are based on feature fusion (b-1). Our proposed Plug-and-play Parameter-intrinsic Watermarking method P2Mark (b-2) flexibly integrates the watermark into the parameters of the generative model, which cannot be removed or manipulated after an open source release. 
\lz{Audio watermarking methods. Existing audio watermarks fall into (a) audio-level watermarking and (b) feature-level watermarking. In the former, watermark embeddings are added to the generated audio; in the latter, watermark embeddings are added to the acoustic features. However, these methods are not applicable in open-source scenarios because users can easily modify the code and remove the watermarks during audio generation. (c) This paper introduces P2Mark, a parameter-level watermarking technique that integrates the watermark into the released model weights, making it impossible for users to delete the watermarks in open-source scenarios. 
% [Comment: Perhaps this figure could be divided into three parts: (a) Audio-level Watermarking, (b) Feature-level Watermarking, and (c) Parameter-level Watermarking.]
}}
\label{fig:type_fig}
\end{figure*}

\lz{

To ensure the secure usage of NSG, audio watermarking has emerged as a critical research topic. Its primary objective is to embed imperceptible watermarks into synthetic audio, enabling the differentiation of synthetic speech without compromising the quality of the synthetic audio. Current audio watermarking methods can be roughly categorized into two types: audio-level watermarking \cite{o2024maskmark,liu2023detecting,liu2023dear,chen2023wavmark,san2024proactive} and feature-level watermarking \cite{liu2024groot,zhou2024traceablespeech,zhou2024wmcodec}. As illustrated in Figure \ref{fig:type_fig}, audio-level watermarking embeds watermark information directly into the synthetic speech, thereby protecting and tracing the speech itself. In contrast, feature-level watermarking treats the watermark as an additional feature, which is encoded and fused with the acoustic features before speech generation. Although these methods present promising solutions for secure NSG usage, they exhibit significant limitations in open-source scenarios where both source codes and model weights are publicly available. In such environments, users can easily modify just a few lines of code to remove the watermark module (for example by extracting the output before the watermarking stage, or remove watermark embedding before feature fusion, as illustrated in Figure \ref{fig:type_fig}(a),(b)), indicating a lack of white-box protection capability. 
Furthermore, these methods fail to protect the copyrights of open-source models and the intellectual property of developers.

}

\lz{

To tackle the challenges of traceability and copyright protection in scenarios where source codes and model weights are open-sourced, we propose a novel parameter-level watermarking technique that embeds watermarks directly into the model weights (as illustrated in Figure \ref{fig:type_fig}(c)). Before implementing this approach, several critical design considerations must be emphasized. Firstly, it is essential to enable publishers to conveniently and flexibly embed watermark information into generative models. This necessitates that the method should be easily integrated into existing frameworks in a plug-and-play manner, allowing for dynamic updates to the watermark information without the need for retraining the watermark model. Secondly, for users with access to the source code and model weights, we must ensure that no backdoor interfaces exist that could permit unauthorized modification of the embedded watermarks. Once the model weights are released, the watermark information must remain immutable.

}

\lz{

To meet the above requirements, inspired by injecting knowledge into models through adapters \cite{wang2021k,hu2021lora}, we propose a parameter-level watermarking technique, namely \textbf{P}lug-and-play \textbf{P}arameter-level Water\textbf{Mark}ing (P2Mark). Specifically, we embed watermark information into the pre-trained speech generation model through a Watermark Low-Rank Adaptation (WM-LoRA) module. 
% This module is lightweight and has lower requirements in terms of computational resources and adaptation time. 
During training, we propose a gradient orthogonal projection optimization strategy to ensure the quality of the generated audio and the accuracy of watermark preservation. 
% After training, publishers can randomly set the watermark information without the need for a retraining process. 
After training, publishers can flexibly set different watermark information at any time without the need for retraining. 
For the release of model weights, the watermark information is merged into the model weights, ensuring that users cannot remove or manipulate the watermark information.

}

% To verify the effectiveness and applicability of the proposed method, since the current mainstream neural speech generation systems primarily generate speech waveforms through two methods: a vocoder to reconstruct waveforms from mel-spectrograms and a codec decoder to convert discrete acoustic tokens into waveforms. Based on our proposed method, we design two implementation schemes: Vocoder-based P2Mark (P2Mark-Vocoder) and Codec-based P2Mark (P2Mark-Codec). This covers waveform generation in mainstream neural speech generation architectures and verifies that P2Mark can protect most neural speech generation models.

% \ry{
% Current mainstream NSG systems primarily generate speech waveforms through two types of waveforms decoders: a vocoder to reconstruct waveforms from mel-spectrograms and a codec decoder to convert discrete acoustic tokens into waveforms. 
% To evaluate the effectiveness and applicability of our proposed method, we implement P2Mark on two mainstream waveforms decoders in NSG: Vocoder-based P2Mark (P2Mark-Vocoder) and Codec-based P2Mark (P2Mark-Codec).
% Experimental results demonstrate that our approach achieves outstanding performance across watermark extraction accuracy, watermark imperceptibility, and robustness against attacks.
% }

\lz{

Currently, NSG systems primarily rely on two types of waveform decoders for speech generation: the vocoder to reconstruct waveforms from mel-spectrograms and the codec decoder to convert discrete acoustic tokens into waveforms. To evaluate the effectiveness of our proposed method, we implement P2Mark on these two types of decoders, making our solution applicable to mainstream NSG systems: Vocoder-based P2Mark (P2Mark-Vocoder) and Codec-based P2Mark (P2Mark-Codec). Experimental results show that our approach achieves outstanding performance in terms of watermark extraction accuracy, watermark imperceptibility, and robustness against attacks.

}

% \lz{

% To evaluate the effectiveness and applicability of our proposed method, we implement P2Mark on two mainstream neural speech generation architectures: Vocoder-based P2Mark (P2Mark-Vocoder) and Codec-based P2Mark (P2Mark-Codec). Experimental results demonstrate that in open-sourced scenarios, our approach achieves outstanding performance across watermark extraction accuracy, watermark imperceptibility, and robustness against attacks. The main contributions of this work are summarized as follows:

% }

The main contributions of this paper can be summarized as follows:
\begin{itemize}

\lz{

\item \textbf{Parameter-level plug-and-play watermarking.} 
We propose P2Mark, a parameter-level watermarking technique that enables seamless integration of watermark information directly into model weights through a plug-and-play module. Even when the model is fully open-sourced (including both source code and model weights), attackers remain unable to remove or tamper with the embedded watermark.

\item \textbf{Optimization strategy.}
To mitigate the interference between watermark optimization and audio quality optimization, we introduce a training optimization method employing gradient orthogonal projection. This approach effectively balances the preservation of audio quality with the accuracy of watermark extraction during training.

\item \textbf{Comprehensive validation.} 
To comprehensively evaluate the effectiveness of our proposed method, we design implementation schemes of our P2Mark for mainstream NSG architectures. Extensive experiments demonstrate our universality across different speech generation frameworks, achieving superior performance in terms of watermark extraction accuracy, watermark imperceptibility, and robustness against various attacks.

}
\end{itemize}

\section{Related Work}

% In this section, we will first introduce the latest NSG methods and the waveform decoders used for NSG. Then, we will review the development and classification of digital watermarking. Finally, we will provide a detailed overview of the related work on audio watermarking.

\lz{

In this section, we first review the latest NSG methods and the waveform decoders employed for NSG. Then, we review the evolution and classification of digital watermarking techniques. Finally, we present a comprehensive overview of existing research on audio watermarking.

}

\subsection{Neural Speech Generation}

\lz{

In recent years, advancements in deep learning and artificial intelligence have significantly enhanced the development of NSG techniques, particularly in terms of speech naturalness, enabling the synthesis of speech that closely rivals real human voices. WaveNet \cite{van2016wavenet} was the first to propose using neural networks to directly generate waveforms from linguistic features. Since then, numerous NSG models have been proposed, among which text-to-speech (TTS) has garnered the most attention. Some classic TTS models such as Tacotron 1/2 \cite{wang2017tacotron, shen2018natural}, FastSpeech 1/2 \cite{ren2019fastspeech, renfastspeech}, GradTTS \cite{popov2021grad}, Glow-TTS \cite{kim2020glow}, and VITS \cite{kim2021conditional} have continuously improved the naturalness of generated speech. With the integration of large language models (LLMs) into speech generation, advanced NSG methods like VALL-E \cite{wang2023neural}, BASE TTS \cite{lajszczak2024base}, Seed-tts \cite{anastassiou2024seed}, Clam-tts \cite{kim2024clam}, Cosyvoice 1/2 \cite{du2024cosyvoice,du2024cosyvoice2} can not only generate speech that is indistinguishable from real human speech but also clone the target voice through target speech prompts. Beyond TTS, related NSG tasks include voice conversion \cite{huang24_interspeech, baade2024neural}, which alters the speaker’s identity while preserving linguistic content, and speech editing \cite{wang2024emotion, peng2024voicecraft}, which enables precise modifications to the content without changing the speaker’s voice. These tasks further expand the scope and capabilities of NSG technologies.

}

\lz{

Among the above NSG frameworks, the waveform decoder plays a critical role in speech generation. Currently, waveform decoders primarily fall into two categories: vocoders and neural codec decoders. Vocoders convert mel-spectrograms into waveforms \cite{van2016wavenet,prenger2019waveglow,kumar2019melgan,kong2020hifi,lee2022bigvgan,song2020bridging}, whereas neural codec decoders encode speech into discrete tokens and reconstruct the waveform \cite{zeghidour2022soundstream,defossezhigh,yang2023hifi,ren2024fewer,kumar2024high,zhang2024speechtokenizer}. To evaluate the effectiveness and universality of P2Mark, we integrate our proposed method into these two types of decoders, resulting in two frameworks P2Mark-Vocoder and P2Mark-Codec.
}

\subsection{Watermarking}
Watermarking methods involve embedding watermarks into cover media in the form of labels, tags, or digital signals \cite{gao2010geometric,patel2011unified,agarwal2019survey}. 
Initially, digital watermarking technology was primarily applied to images to prevent illegal copying and distribution of digital images \cite{rahman2013dwt,zhang2019robust}.
With the advancements in generative artificial intelligence models, the application scenarios and functions of digital watermarking technology have been significantly expanded.
In terms of carriers, watermarking methods are not limited to images but are also widely applied to various multimedia content such as audio \cite{chen2023wavmark} and video \cite{zhou2022robust}. 
In terms of functionality, watermarking technology is not only used to protect the copyrights of authentic multimedia content but is also gradually being applied to the marking of synthetic content \cite{ren2024copyright,juvela2025audio} and the protection of intellectual property rights related to generative models \cite{feng2024aqualora}. 

% Based on the method of watermark embedding, mainstream watermarking methods can be classified into two types: Post-hoc watermarking and generative model watermarking. 
% Post-hoc watermarking involves embedding watermark information into the generated data after multimedia content has been created and can be viewed as a form of data watermarking\cite{rahman2013dwt,zhang2019robust}.
% Generative model watermarking integrates the watermarking process with content generation, utilizing the generative model to accomplish both tasks simultaneously\cite{wen2023tree}. This paper focuses on audio watermarking. Therefore, the next subsection will provide a detailed overview of the related work on audio watermarking.

\subsection{Audio Watermarking}

% Audio watermarking was initially developed to embed copyright or other information into audio signals for the purpose of marking and protecting audio content \cite{boney1996digital}. With the advancement of NSG technologies, there has been a growing demand for tracing synthetic speech and protecting the intellectual property of NSG models.

\lz{

Audio watermarking was initially developed to embed copyright or other information into audio signals for the purpose of marking and protecting audio content \cite{boney1996digital}. With the advancement of NSG technologies, there has been a growing demand for tracing synthetic speech and protecting the intellectual property of NSG models. Current audio watermarking techniques can be broadly divided into three categories: audio-level watermarking, feature-level watermarking, and parameter-level watermarking.

}

In traditional digital audio steganography and audio copyright protection, the target of watermark embedding is the audio signal itself. Therefore, these approaches typically embed watermarks directly into the audio and are often referred to as post-hoc audio watermarking methods. Based on the embedding carrier and embedding manner, we refer to this class of watermarking as audio-level watermarking, as illustrated in Figure \ref{fig:type_fig}(a).
Audio-level watermarking has undergone extensive development over the years.
Early audio watermarking methods primarily embedded watermarks in the time or frequency domain using manual techniques \cite{zhang2019robust,zhang2020time,hu2020selection,qin2023lattice}, but these methods affected audio quality and had poor robustness. Some end-to-end Post-hoc audio watermarking methods based on deep neural networks have achieved more powerful performance \cite{o2024maskmark,liu2023detecting,liu2023dear,chen2023wavmark,san2024proactive}, including better imperceptibility, watermark capacity, and robustness. 
Maskmark \cite{o2024maskmark} embeds secret watermarks in audio through multiplicative spectral masking to enhance robustness. Timbre \cite{liu2023detecting} embeds watermarks in the frequency domain, adopting a repetitive embedding strategy to further enhance robustness. DeAR \cite{liu2023dear} designs a watermarking framework based on deep learning and has developed a distortion layer to defend against audio re-recording attacks. WavMark \cite{chen2023wavmark} designs an advanced watermark embedding and detection framework, improving the capacity and robustness of Post-hoc watermarks. AudioSeal \cite{san2024proactive} designs a watermark embedding and detection framework specifically for local detection of AI-generated speech, achieving high accuracy and robustness.
% However, because the generation process and the watermarking process are two independent stages in Post-hoc audio watermarking, once the code and model weights of the generation model are open-sourced, users can choose to skip the watermarking process, circumventing the addition of watermarks. 

Although audio-level watermarking offers high watermark capacity, when applied to the identification of AI-generated speech, the cascaded system introduces cumulative errors and increases processing time. To further enhance the imperceptibility of generated speech and reduce the time overhead of the watermarking process, several audio watermarking methods specifically designed for proactive tracing of synthetic speech have been proposed. These approaches integrate the watermark into the generation model's input features by fusion of the watermark features obtained after a watermark encoder. In this way, the watermark embedding process is unified with speech generation. We refer to this type of method as feature-level watermarking, as illustrated in Figure \ref{fig:type_fig}(b).
GROOT \cite{liu2024groot} adds watermarks to the initial diffusion noise of the diffusion model. TraceableSpeech \cite{zhou2024traceablespeech} concatenates the watermarked features and quantized acoustic tokens, then feeds them into the Decoder of the Codec. WMCodec \cite{zhou2024wmcodec} integrates pre-quantization acoustic tokens with watermark features through an attention mechanism and then feeds them into the Decoder of the Codec after quantization.

End-to-end training enables audio-level watermarking to achieve faster processing speeds and better watermark imperceptibility, making it suitable for deployment in NSG APIs. However, with the rapid growth of the open-source generative AI community, an increasing number of NSG models and their source codes have been released publicly. In this context, neither audio-level watermarking nor feature-level watermarking can effectively protect open-source NSG models: for the former, the post-hoc watermarking module can be bypassed by directly accessing the intermediate audio output; for the latter, the watermark feature can be easily altered by modifying just a few lines of code. To prevent the misuse of open-source NSG models and to protect their copyrights, it becomes necessary to embed the watermark directly into the model parameters, making the watermark inseparable from the model itself. We refer to this type of method as parameter-level watermarking, as illustrated in \ref{fig:type_fig}(c).
There are two concurrent explorations of model-parameter-level audio watermarking methods.
San Roman et al. \cite{san2024latent} proposed an audio watermarking method that embeds an audio-level watermark into the training data, such that the NSG model learns to fit the distribution of watermarked audio and consequently generates audio containing detectable watermarks. 
% To ensure that the generation process does not degrade the embedded watermark, it is necessary to simulate attacks on the codec used by the NSG model during training to enhance watermark robustness.
However, this approach has several limitations: first, it requires watermarking the entire training corpus, which can lead to a decline in training data quality; second, it may be difficult to support versioning of large models or to adapt already-trained models; and finally, the robustness enhancement must be specifically tailored to the neural speech generation method being used.
The most closely related work to ours is HiFiGANw \cite{cheng2024hifi}, which fine-tunes the neural vocoder HiFi-GAN using a combination of watermark extraction loss and speech quality loss.
However, in HiFiGANw, the embedded watermark is fixed during fine-tuning, and changing the watermark requires re-finetuning the model from scratch, limiting its practicality and flexibility in real-world applications. 
\ry{
To overcome the aforementioned limitations, we propose a novel plug-and-play parameter-level watermarking technique. Instead of fine-tuning the NSG model with a fixed watermark during training, we freeze the NSG parameters and fine-tune a specially designed watermark adapter, through which the watermark information is embedded. Consequently, rather than using a fixed watermark during training, random watermark information is generated dynamically. After training, watermark information can be flexibly merged into the NSG model parameters through the adapter, allowing different watermarks to be embedded without retraining. When releasing the model, only the merged version needs to be published, thereby ensuring both flexibility before release and security after release.
}
% To overcome the above limitations, we propose a novel parameter-level watermarking technique,
% \lz{
% [Comment: To overcome the above limitations, we propose a novel parameter-level watermarking technique, xxx] 
% }

\begin{figure*}[t]
\centering
  \includegraphics[width=\linewidth]{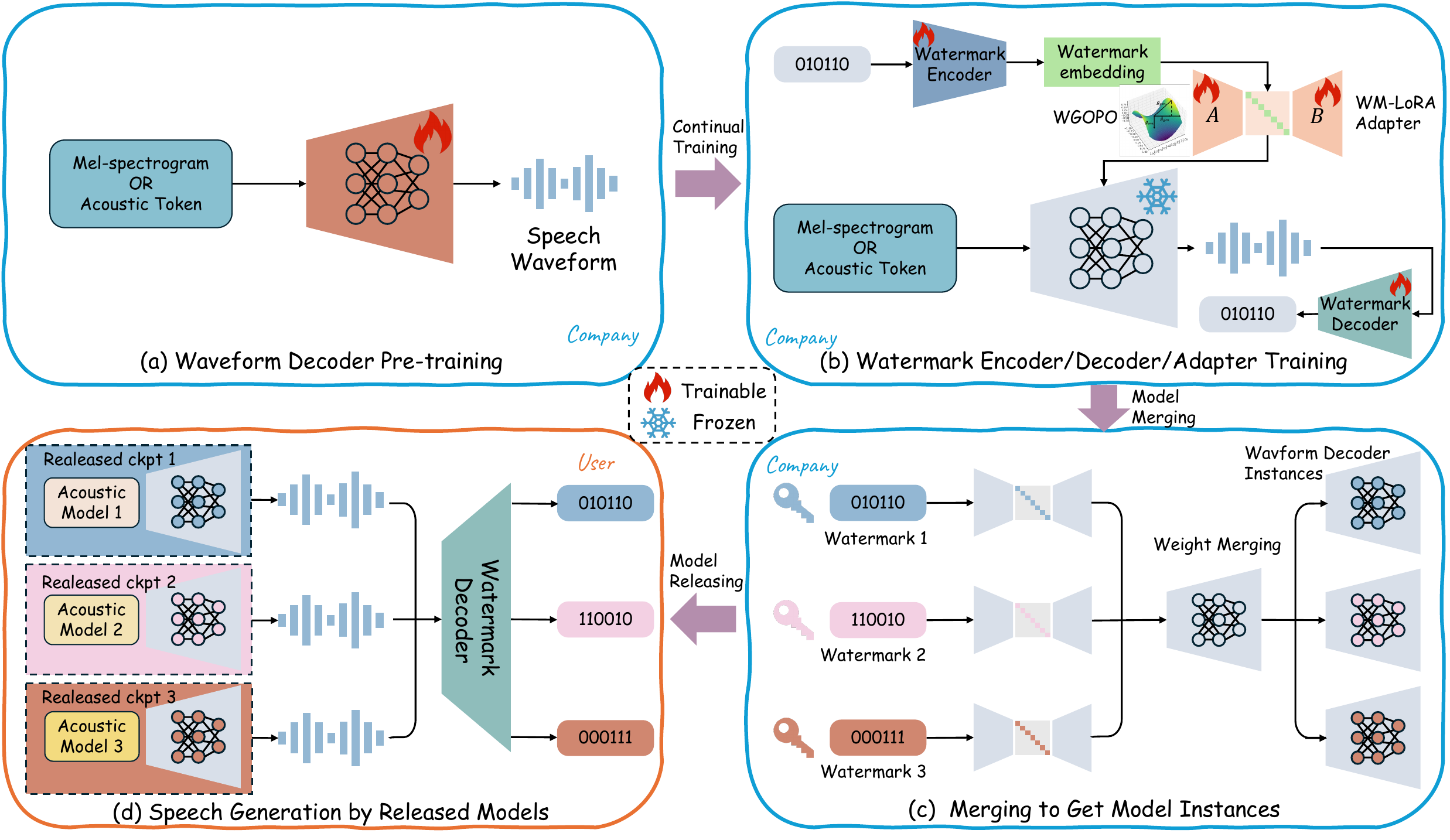} 
\caption{The overall framework of the proposed method. First, pre-train the waveform decoder (part a), then train the watermark encoder, watermark decoder, and watermark adapter (part b). Subsequently, different watermarks can be merged with the weights of the adapter and waveform decoder to obtain different instances of models (part c). Finally, these instances of models are released as part of the speech generation models. The speech generated by the open-source model can be detected by the watermark decoder (part d).}
\label{fig:main_fig}
\end{figure*}

\section{Proposed Method}

\subsection{Overview of P2Mark}

Figure \ref{fig:main_fig} provides an overview of our proposed P2Mark method. Our approach enables plug-and-play parameter-level watermark fusion for NSG, providing white-box protection for open-source models. 

Firstly, as shown in Figure \ref{fig:main_fig}(a), we need to pre-train a speech waveform decoder that can convert acoustic features obtained from the acoustic model (such as mel-spectrograms or acoustic tokens) into speech waveforms. Then, as illustrated in Figure \ref{fig:main_fig}(b), we train a Plug-and-play WM-LoRA module. The watermark, after being encoded by the watermark encoder, results in a watermark embedding. This embedding is then combined with WM-LoRA to form an adapter for the pre-trained waveform decoder and is trained together with the watermark encoder module and the watermark decoder module. During this process, the parameters of the original waveform decoder are frozen. While training the adapter, we employ the Watermarking Gradient Orthogonal Projection Optimization (WGOPO) method to minimize the conflict between watermark optimization and generation optimization. 
After successfully training the watermark encoder, the watermark adapter module, and the watermark decoder, as shown in Figure \ref{fig:main_fig}(c), we can integrate different watermark embeddings into the weights of the pre-trained waveform decoder through the watermark adapter merging.
This results in various instances of the waveform generator containing different watermark information. Finally, as shown in Figure \ref{fig:main_fig}(d), we release the instances of the waveform generator containing watermark information as part of the speech generation model. The speech generated by these open-source speech generation models can be accurately traced by our watermark decoder. 

\begin{figure*}[t]
\centering
  \includegraphics[width=\linewidth]{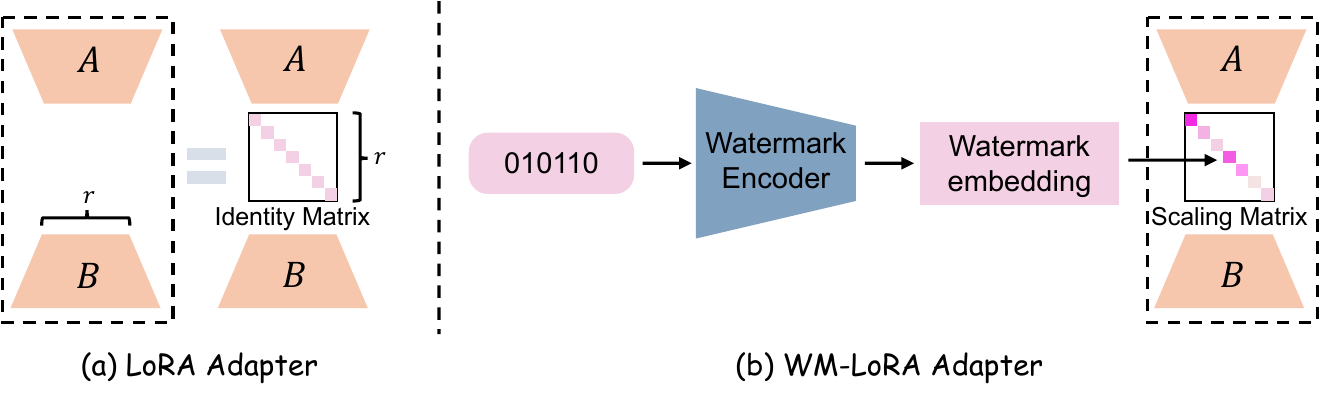} 
\caption{The LoRA Adapter (a) and the Plug-and-play Watermarking Adapter, WM-LoRA Adapter (b).}
\label{fig:lora_fig}
\end{figure*}

% \lz{
% Comment: Perhaps reorganize the paper to emphasize the proposed method.

% %% Part1: how to realize P2Mark
% \subsection{P2Mark}
% % Target xxx

% \subsubsection{Watermark Encoder and Decoder}
% % For watermark preservation

% \subsubsection{Plug-and-play Watermarking Adapter}
% % For plug-and-play adaption

% \subsubsection{Loss Function}

% %% Part2: how to implement P2Mark into NSG
% \subsection{Incorporating P2Mark into NSG Frameworks}

% \subsubsection{P2Mark-Vocoder}

% \subsubsection{P2Mark-Codec}

% \subsubsection{Loss Function}
% % P2Mark loss + NSG Loss

% %% Part3: How to Optimize the Joint Framework
% \subsection{Optimization Strategy}
% % Since there is a conflict between different losses, how to solve this problem?
% }

% \ry{
% \subsection{Overview of P2Mark}
% \subsection{Plug-and-play Watermarking Adapter}
% \subsection{Incorporating P2Mark into NSG Frameworks}
% \subsubsection{P2Mark-Vocoder}
% \subsubsection{P2Mark-Codec}
% \subsection{Parameter-level Watermark Fusion}
% \subsubsection{Parameter-level Watermark Fusion Training}
% \subsubsection{Watermarking Gradient Orthogonal Projection Optimization}
% }

\subsection{Plug-and-play Watermarking Adapter}
\label{sec_ppwm}

The purpose of digital watermarking is to embed extractable information into the target object to be marked. For practical deployment, flexibility in watermarking is also a crucial factor, alongside security and reusability. First, regarding security, if the watermark remains immutable after model training, there is a risk of watermark information leakage. Thus, allowing the watermark embedded in the model parameters to be modified at any time before release, without the need for retraining, is a critical consideration for making parameter-level watermarking practical. Second, in terms of reusability, embedding different watermark information into the waveform generation module of neural speech generation models can produce distinct model instances, enabling effective version control for speech generation systems. Without considering such flexibility, a straightforward approach is to jointly fine-tune the speech generation model and the watermark extractor, as done in HiFiGANw \cite{cheng2024hifi}. However, this leads to a significant limitation: once the model is trained, the embedded watermark becomes fixed, and updating the watermark requires retraining the entire model. In this work, we aim to develop a watermarking method that can flexibly modify the watermark information embedded in model parameters without retraining. To address these requirements, we propose a plug-and-play audio watermark adapter method. With only a one-time fine-tuning of the watermark adapter, subsequent watermark information can be freely selected and merged into the model parameters through adapter merging, enabling flexible and efficient watermark integration.

% The purpose of watermarking is to embed extractable information into a model or data. Flexibility is a crucial aspect of the watermarking method. If flexibility is not considered, to achieve white-box protection when the codes and model weights of the speech generative model are open-sourced, we can directly fine-tune the speech generation model with the watermark extractor together. However, this approach is impractical in real-world scenarios, as it is needed to retrain the model each time the watermark information needs updating. We hope that the watermarking method can \textbf{flexibly change the watermark information embedded in the model parameters without re-training}. To address these requirements, we propose a plug-and-play watermarking adapter method for audio watermarking. After a single training session, we can freely choose when and what content of watermark information to be integrated with the parameters by merging the adapter.

LoRA\cite{hu2021lora} is a technique used to efficiently fine-tune LLMs by adapting two low-rank matrices with a small number of parameters, thereby reducing computational costs and memory usage while maintaining performance. Additionally, LoRA can be considered a plug-and-play module that can be merged with the original parameters of the model at any time.
Layers of neural networks can perform matrix multiplication, as shown in Figure \ref{fig:lora_fig}(a). For a pre-trained weight matrix $W_0 \in \mathbb{R}^{d \times k}$, forward pass $h = W_0x$ modified by original LoRA yields:
\begin{equation}
    h = W_0x + \frac{\alpha}{r}\Delta Wx
      = W_0x + \frac{\alpha}{r}BAx,
\end{equation}
where matrix $ B \in \mathbb{R}^{d \times r} $, $ A \in \mathbb{R}^{r \times k} $, $\Delta W = BA$, rank $r \ll min(d, k)$, and $ \alpha $ is a scaling factor. During training, $W_0$ is frozen and does not receive gradient updates, while $A$ and $B$ contain trainable parameters.

As shown in Figure \ref{fig:lora_fig}(b), we insert a diagonal matrix $ S $ between matrices $ B $ and $ A $ as a scaling matrix, replacing the scaling factor $\frac{\alpha}{r}$, thus modifying the LoRA update formula to: 
\begin{equation}
    h = W_0x + BSAx.
\end{equation}
This scaling matrix allows us to integrate variable watermark information into the target speech generation model.

We employ learnable embeddings as a watermark encoder $E_{wm}$ to convert a watermark message of length $l$ into an embedding of length $r$. 
For the $i$-th bit $w_i$ of a binary watermark $w$, we obtain an embedding vector $ emb_i \in \mathbb{R}^r $ for each position $ i $ with binary states 1 through the embedding layer. 
Thus, the mapping function can be expressed as follows:
\begin{equation}
    E_{wm}^i(w_i) = 
    \begin{cases}
        emb_i, & \text{if } w_i = 1, \\
        \mathbf{0}, & \text{otherwise}.
    \end{cases}
\end{equation}
For a given binary watermark \( W = \{w_0, w_1, \ldots, w_l\} \), the scaling matrix \( S \) is constructed as:
\begin{equation}
    S = \mathrm{diag}\left( \textbf{1} + \frac{1}{\sqrt{l}} \sum_{i=1}^{l} E_{wm}^i(w_i) \right).
\end{equation}
We inject the modified LoRA adapter into the convolution layer in the pre-trained speech waveform decoder model.
In each iteration of the training, we use a batch of random binary watermark messages.
Once the matrices $A$ and $B$ are trained, we can use the watermark encoder $E_{wm}$ to obtain the scaling matrix $S$, and then merge it into the model weights by computing $ W_{\text{watermarked}} = W_0 + BSA $. 
Thus, we achieve plug-and-play watermarking for speech generation models, enabling the generation of the model checkpoint instances with specific watermarks at any time.

\begin{figure*}[t]
\centering
  \includegraphics[width=\linewidth]{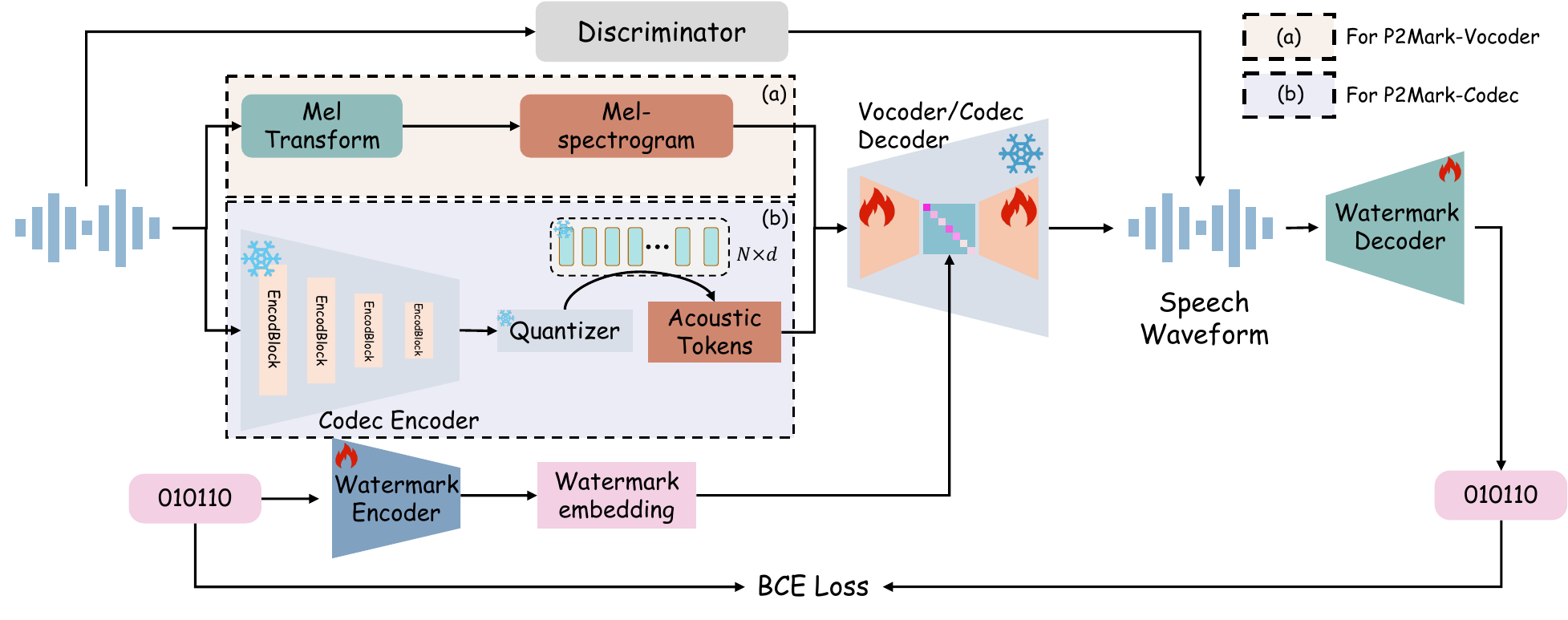} 
\caption{
\ry{The framework of P2Mark-Vocoder and P2Mark-Codec. It consists of three modules: a watermark encoder, a watermark adapter and a watermark decoder. To evaluate the generalizability of our method, we apply P2Mark to two mainstream waveform decoders, vocoder and codec decoder, resulting in P2Mark-Vocoder and P2Mark-Codec. The primary difference between the two lies in the type of acoustic features they use: P2Mark-Vocoder operates on mel-spectrograms, whereas P2Mark-Codec uses acoustic tokens. Therefore, P2Mark-Vocoder includes an additional mel-spectrogram extraction module (a), while P2Mark-Codec incorporates a codec encoder and a quantizer (b).}
}
\label{fig:vocoder_codec_fig}
\end{figure*}

% \subsection{P2Mark-Vocoder and P2Mark-Codec}
\subsection{Incorporating P2Mark into NSG Frameworks}

% Since Vocoder and Codec Decoder are waveform decoders used in most existing NSG methods, we designed two types of waveform decoders with P2Mark, referred to as P2Mark-Vocoder and P2Mark-Codec, As illustrated in Figure \ref{fig:vocoder_codec_fig}.

\lz{

Existing NSG techniques primarily employ two types of waveform decoders: vocoders and codec decoders. To evaluate the universality of our proposed method, we integrate P2Mark into these waveform decoders, resulting in P2Mark-Vocoder and P2Mark-Codec. The detailed framework is illustrated in Figure \ref{fig:vocoder_codec_fig}. For implementation purposes, we select two representative vocoders and codec decoders, HiFi-GAN and HiFi-Codec, for our experiments and analysis. 
% Notably, our P2Mark is also compatible with other vocoders and codec decoders.

}

\subsubsection{P2Mark-Vocoder}

% \begin{figure*}[t]
% \centering
%   \includegraphics[width=\linewidth]{figs/vocoder.pdf} 
% \caption{The framework of P2Mark-Vocoder. P2Mark-Vocoder consists of a Generator, a Discriminator, a Watermark Encoder module, and a Watermark Decoder module.}
% \label{fig:vocoder_fig}
% \end{figure*}

The target of the vocoder is to convert mel-spectrograms into waveforms. GAN-based vocoders are superior in inference speed and synthesis quality when reconstructing an audible waveform from mel-spectrograms. HiFi-GAN is a typical representative of GAN-based vocoders and has been widely used in various speech and audio generation methods\cite{ren2019fastspeech,popov2021grad,casanova2022yourtts}. Therefore, we chose HiFi-GAN as the base model and combined it with P2Mark to propose the parameter-level watermark vocoder, P2Mark-Vocoder. Other vocoders can adopt a similar method to add watermarks. P2Mark-Vocoder consists of four components: a generator, a discriminator, a watermark encoder module, and a watermark decoder module. The framework of P2Mark-Vocoder is as shown in Figure \ref{fig:vocoder_codec_fig} (a).

\textbf{Generator and Discriminator.}
The generator and discriminator of P2Mark-Vocoder are the same as HiFi-GAN. The generator is a fully convolutional neural network that takes a mel-spectrogram as input and outputs a sequence with the same time resolution as the original waveform.  
The discriminator includes a multi-period discriminator (MPD) and a multi-scale discriminator (MSD). 

\textbf{Watermark Encoder Module and Watermark Decoder Module.}
The watermark encoder module serves to encode the input watermark information into a fixed-length embedding, which is subsequently integrated into the model as a component of the P2Mark. This module is composed of an embedding layer that employs orthogonal initialization and weight normalization. The watermark decoder module is composed of a ResNet\cite{he2016deep} and a linear classification layer, and its purpose is to extract features from the mel-spectrogram to obtain the watermark information.

\subsubsection{P2Mark-Codec}

% \begin{figure*}[t]
% \centering
%   \includegraphics[width=\linewidth]{figs/codec.pdf} 
% \caption{The framework of P2Mark-Codec. P2Mark-Codec consists of a Codec Encoder, a Quantizer, a Codec Decoder, a Discriminator, a Watermark Encoder module, and a Watermark Decoder module}
% \label{fig:codec_fig}
% \end{figure*}

The neural audio codec was initially proposed as a method for compressing audio. It converts audio into a compressed representation, which can then be restored to an audio waveform through a codec decoder. With the impressive performance of LLMs in generating various modalities, a new class of speech generation methods based on LLMs has emerged. These methods utilize acoustic tokens obtained from neural audio codecs as the acoustic representation of audio. The language model is then employed to predict these acoustic tokens, which are subsequently decoded into waveforms using the codec decoder. HiFi-Codec\cite{yang2023hifi} is a well-performing codec method. HiFi-Codec achieves superior reconstruction performance compared to the classic codec method EnCodec \cite{defossezhigh} with the same number of codebooks\cite{wu2024codec}.
Therefore, we chose HiFi-Codec as the base model and combined it with P2Mark to propose the parameter-level watermark codec, P2Mark-Codec. 
Other codecs can also easily implement watermark addition using a similar approach.
P2Mark-Codec consists of six components: a codec encoder, a quantizer, a codec decoder, a discriminator, a watermark embedding module, and a watermark extraction module. The framework of P2Mark-Codec is as shown in Figure \ref{fig:vocoder_codec_fig} (b).

\textbf{Encoder, Decoder and Discriminator.}
The encoder is a fully convolutional neural network. The input passes through a one-dimensional convolution, followed by four convolutional blocks. 
The decoder uses a structure that mirrors the encoder.  
The discriminator includes the MPD and MSD from HiFi-GAN, as well as the multi-scale STFT discriminator (MS-STFTD) from EnCodec. 

\textbf{Quantizer.}
We use residual vector quantization to quantize the output of the encoder and learn $ N_q $ sets of codebooks. The unquantized output of the encoder is quantized by the first layer of the learnable codebook, and the quantization residual is calculated. Then, the residual is iteratively quantized through a series of additional $ N_q-1 $ vector quantizers. 

\textbf{Watermark Encoder and Decoder Module.}
The watermark encoder and decoder modules are identical to those used in P2Mark-Vocoder.

Both P2Mark-Vocoder and P2Mark-Codec adopt a GAN-based structure. The key difference lies in the input representation: P2Mark-Vocoder first converts the waveform into a mel-spectrogram, which is then reconstructed by the generator, while P2Mark-Codec encodes the waveform into discrete acoustic tokens through a codec encoder and quantizer, and subsequently reconstructs it using the generator. In both P2Mark-Vocoder and P2Mark-Codec, a watermark adapter is introduced and fine-tuned on top of a pretrained model. Therefore, the encoder, quantizer, and generator of the pretrained model are frozen during training, and only the generator, discriminator, and watermark encoding and extraction modules are trainable.

\subsection{Parameter-level Watermark Fusion}
To achieve open-source white-box protection for NSG models, we integrate the watermark information into the parameters of the generative model.
The plug-and-play watermark adapter we proposed in session \ref{sec_ppwm} can be fine-tuned on a pre-trained NSG model with frozen parameters. 
This approach avoids the need for training from scratch and ensures that the quality of speech generation does not catastrophically decline due to the embedding of the watermark.
To better achieve efficient fusion of watermark information at the parameter level with the generative model, inspired by the continual learning method Averaged Gradient Episodic Memory (AGEM)\cite{chaudhryefficient}, we propose a novel optimization method, WGOPO. 
In \ref{piwft}, we first introduce the training process of Parameter-level Watermark Fusion, and then in \ref{sec_wgopo}, we will provide a detailed introduction to the WGOPO method we proposed.

\subsubsection{Parameter-level Watermark Fusion Training}
\label{piwft}
Parameter-level watermark fusion training requires simultaneous training of the watermark encoder, the watermark decoder, the discriminator, and the watermark adapter. The loss function includes three parts: the Watermark Loss $\mathcal{L}_{WM}$, the Discriminator Loss $\mathcal{L}_{D}$, and the Generator Loss $\mathcal{L}_{G}$. During the training process, $\mathcal{L}_{WM}$, $\mathcal{L}_{D}$, and $\mathcal{L}_{G}$ are optimized alternately.

\textbf{Watermarking Loss $\mathcal{L}_{WM}$:}
Watermarking Loss is defined as the binary cross-entropy between the output of the watermark extractor and the binary watermark ground truth:
\begin{equation}
    \mathcal{L}_{WM} = - \sum_{i=1}^l w_i  \log \hat{w}_i + (1 - w_i)  \log(1 - \hat{w}_i),
\end{equation}
where $l$ is the length of the watermark sequence, $w_i$ is the ground truth binary watermark, and $\hat{w}_i$ is the predicted binary watermark.

\textbf{Discriminator Loss $\mathcal{L}_{D}$:}
\begin{align}
    \mathcal{L}_{D} &= \mathcal{L}_{Adv}(D; G).
    \label{final_loss_disc} 
\end{align}

\textbf{Generative Loss $\mathcal{L}_{G}$:}
The Generation Loss is composed of three weighted parts: 
(1) an adversarial loss $\mathcal{L}_{adv}$ to enhance the perceptual quality of generated audio;
(2) a feature matching loss $\mathcal{L}_{FM}$ to stabilize GAN training by aligning intermediate discriminator features between real and generated samples;
and (3) a mel-spectrogram reconstruction loss $\mathcal{L}_{Mel}$ to improve training efficiency and spectral fidelity \cite{kong2020hifi,yang2023hifi}.
% the GAN Loss $\mathcal{L}_{adv}$, the Feature Matching Loss $\mathcal{L}_{FM}$, and the Mel-spectrogram Loss $\mathcal{L}_{Mel}$.
\begin{align}
    \mathcal{L}_{G} &= \mathcal{L}_{Adv}(G; D) + \lambda_{fm}\mathcal{L}_{FM}(G; D) + \lambda_{mel}\mathcal{L}_{Mel}(G).
    \label{final_loss_gen}
\end{align}

\begin{algorithm}[h!]
    \caption{Training Strategy for P2Mark}
    \label{alg-1}
    \KwIn{Pre-trained Vocoder/Codec model $\theta$, WM-LoRA $\delta \theta$, Watermark Encoder $E_{wm}$, Watermark Decoder $D_{wm}$, An audio dataset $\mathcal{D}$}
    \KwOut{Fine-tuned WM-LoRA $\delta \theta$}
    \BlankLine
    
    \SetKwProg{Fn}{Function}{:}{}
    \SetKwFunction{FSuba}{Train}
    \SetKwFunction{FSubb}{Generator}
    
    \Fn{\FSuba{$\mathcal{D}$}}{
        Load Pre-trained Vocoder/Codec model $\theta$;
        
        Replace the original Conv1D layers in the generator $\mathcal{G}$ with the Conv1D layers WM-LoRA (Rank = $r$) to get $\mathcal{G}_{LoRA}$;
        
        Freeze parameters in $\mathcal{G}_{LoRA}$ except for the LoRA parameters;
        
        Split $\mathcal{D}$ into batches $\mathcal{D} = \{\mathcal{X}_i\}_{i=1}^{N_b}$;
        
        \For{$i = 1, \cdots, N_b$}{
            
            Randomly generate $K$ bits binary watermark $w$;
            
            Encode $w$ into an embedding $E_{wm}(w)$ of length $r$;
            
            \If{Method is P2Mark-Vocoder}{
                $z_i$ = STFT($\mathcal{X}_i$);
            }
            \ElseIf{Method is P2Mark-Codec}{
                $z_i$ = VQ(Encoder($\mathcal{X}_i$));
            }
            
            $\hat{z}_i$ = $\mathcal{G}_{LoRA}(z_i, E_{wm}(w))$;
            
            Optimize the Discriminator by Discriminator Loss \textcolor[rgb]{0.93,0.0,0.47}{$\mathcal{L}_{D}$};
            
            Calculate the gradient $g_{wm}$ of the generator backpropagated by watermark loss \textcolor[rgb]{0.93,0.0,0.47}{$\mathcal{L}_{WM}$};
            
            Optimize the Generator $\mathcal{G}_{LoRA}$, Watermark Encoder $E_{wm}$, and Watermark Decoder $D_{wm}$;
            
            Calculate the gradient $g_{gen}$ of the generator backpropagated by Generative loss \textcolor[rgb]{0.93,0.0,0.47}{$\mathcal{L}_{G}$};
            
            Project the gradient $g_{gen}$ using WGOPO:
            
            \If{$g_{wm}^\top g_{gen} < 0$}{
                $\tilde{g}_{gen} = g_{gen} - \frac{g_{gen}^\top g_{wm}}{g_{wm}^\top g_{wm}} g_{wm}$;
            }
            
            Optimize the Generator $\mathcal{G}_{LoRA}$;
        }
    }
\end{algorithm}

\begin{figure*}[t]
\centering
  \includegraphics[width=0.9\linewidth]{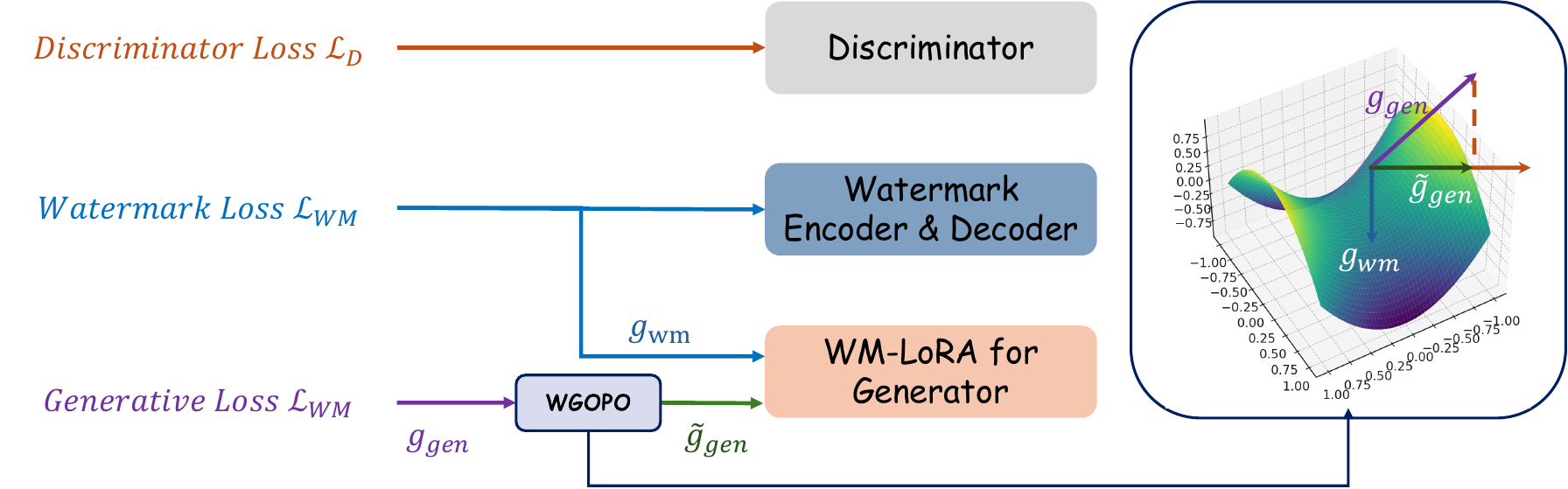} 
\caption{
Training Optimization Process.
The Discriminator Loss $\mathcal{L}_{D}$, Watermark Loss $\mathcal{L}_{wm}$, and Generative Loss $\mathcal{L}_{G}$ are optimized in an alternating manner. 
Both $\mathcal{L}_{wm}$ and $\mathcal{L}_{G}$ jointly update the Wm-LoRA parameters in the generator, which may lead to conflicting optimization directions. 
To address this, we propose WGOPO. When the angle between the gradient $g_{\text{gen}}$ (from the generative loss) and the gradient $g_{\text{wm}}$ (from the watermark loss) exceeds $\pi/2$, $g_{\text{gen}}$ is projected onto the plane orthogonal to the saved $g_{\text{wm}}$, ensuring that the watermark loss does not increase during optimization.
% Schematic diagram of watermarking gradient orthogonal projection optimization. When the angle between the gradient $g_{gen}$ from the generation loss and the gradient $g_{wm}$ from the watermark loss exceeds $\pi/2$, $g_{gen}$ is projected onto a plane orthogonal to saved $g_{wm}$ to ensure that the watermark loss does not increase. 
% \lz{Comment: It would be better to use darker colors for the left parts, and the blue line should be more clearly separated from the grey box.}
}
\label{fig:wgopo_fig}
\end{figure*}

\subsubsection{Watermarking Gradient Orthogonal Projection Optimization}
\label{sec_wgopo}
% We hope that the process of watermarking does not severely affect the performance of the speech generation model. 
% However, optimizing the watermarking process and optimizing the speech generation model's performance are antagonistic. 
% The optimization of the watermarking inevitably affects the performance of the speech generation model. 
% During the joint training of the watermark encoder, the watermark decoder, the watermark adapter, and the generator, the optimization objectives of the watermarking and the generation are significantly different. The optimization directions of the Watermark Loss $\mathcal{L}_{WM}$ and Generative Loss $\mathcal{L}_{G}$ for the generator module are inconsistent. 

We aim for the watermark embedding process to have minimal impact on the performance of the speech generation model. However, the optimization objectives of watermarking and speech generation are inherently misaligned. Since both the watermark loss and the generation loss jointly update the generator, the optimization of the watermark can interfere with the performance of the speech generation model. During the joint training of the watermark encoder, watermark decoder, watermark adapter, and generator, this conflict between the objectives can make it difficult for the generator to converge to a good local optimum. 
% Specifically, the optimization directions of the generator's watermark loss $\mathcal{L}_{WM}$ and generation loss $\mathcal{L}_{G}$ are inconsistent.

To address this challenge, we draw inspiration from the continual learning method AGEM\cite{chaudhryefficient} and propose WGOPO, a strategy that simultaneously optimizes watermark extraction accuracy and waveform generation quality, as illustrated in Figure \ref{fig:wgopo_fig}. Specifically, when the gradient from optimizing $\mathcal{L}_{WM}$ is backpropagated, WGOPO stores the corresponding generator gradient. During subsequent optimization with respect to $\mathcal{L}_{G}$, WGOPO calculates the angle between the current gradient and the stored gradient. If the angle exceeds $\pi/2$ radians, indicating a conflicting optimization direction, WGOPO projects the current gradient onto a direction that ensures $\mathcal{L}_{WM}$ does not increase, thereby mitigating interference between the two objectives.

% Inspired by the continual learning method AGEM\cite{chaudhryefficient}, we propose WGOPO, which simultaneously optimizes both watermark extraction accuracy and waveform generation quality, as shown in Figure \ref{fig:wgopo_fig}. When the gradient from optimizing $\mathcal{L}_{WM}$ is backpropagated, WGOPO stores the gradient that optimizes the generator. During the optimization of $\mathcal{L}_{G}$ for the generator, we calculate the angle between the current gradient direction and the stored gradient direction. If the angle exceeds $\pi/2$ radians, WGOPO projects the current gradient direction to ensure that $\mathcal{L}_{WM}$ does not increase.
The corresponding optimization problem is formulated as:

\begin{equation}
    minimize_{\tilde{g}_{gen}} \frac{1}{2} \lVert g_{gen} - \tilde{g}_{gen} \rVert _2^2 \quad s.t. \quad \tilde{g}_{gen}^\top g_{wm} \ge 0.
\end{equation}

when the gradient $g_{gen}$ violates the constraint, it is projected via:
\begin{equation}
    \tilde{g}_{gen} = g_{gen} - \frac{g_{gen}^\top g_{wm}}{g_{wm}^\top g_{wm}} g_{wm},
\end{equation}
where $g_{gen}$ is the current gradient, $g_{wm}$ is the saved gradient from the $\mathcal{L}_{WM}$ optimization, and $\tilde{g}_{gen}$ is the projected gradient that satisfies the constraint. 
% This ensures that the optimization of the generator by the $\mathcal{L}_{G}$ does not increase the $\mathcal{L}_{WM}$. 
The parameter-level watermark fusion training algorithm incorporated the WGOPO is shown as Algorithm \ref{alg-1}.

\section{Experiments}

\subsection{Experimental Settings}

\subsubsection{Datasets}

% We conducted experiments using dataset LibriTTS\cite{zen2019libritts}, a widely recognized multi-speaker English corpus derived from the LibriVox project's audiobooks. LibriTTS is specifically designed for TTS applications, featuring high-quality multi-speaker voice data. It includes approximately 585 hours of voice data sampled at 24kHz from 2,456 speakers. 
% For our training data, we employed the subsets train-clean-100, train-clean-360, and train-other-500. For validation purposes, we randomly selected 200 samples from the dev-clean and dev-other subsets. For our testing set, we randomly extracted 1,000 samples, each ranging from 1 to 10 seconds in duration, from the test-clean and test-other subsets.

We conducted experiments using the LibriTTS dataset\cite{zen2019libritts}, a widely recognized multi-speaker English corpus derived from audiobooks of the LibriVox project. LibriTTS is specifically designed for text-to-speech (TTS) applications, offering high-quality, multi-speaker audio data. It contains approximately 585 hours of recordings sampled at 24kHz from 2,456 speakers.
For training, we utilized the train-clean-100, train-clean-360, and train-other-500 subsets. For validation, we randomly selected 200 samples from the dev-clean and dev-other subsets. For testing, we randomly extracted 1,000 samples from the test-clean and test-other subsets, with each sample ranging from 1 to 10 seconds in duration.
In both the training and testing phases, we embedded randomly generated $n$-bit binary sequences as watermarks, ensuring that the watermark information varied across different samples.

\subsubsection{Baselines}

We compare P2Mark against two audio-level audio watermarking methods WavMark\footnote{https://github.com/wavmark/wavmark}\cite{chen2023wavmark} and AudioSeal\footnote{https://github.com/facebookresearch/audioseal}\cite{san2024proactive}, and two feature-level audio watermarking methods TraceableSpeech\cite{zhou2024traceablespeech}\footnote{https://github.com/zjzser/TraceableSpeech} and WMCodec\cite{zhou2024wmcodec}\footnote{https://github.com/zjzser/WMCodec}.

\begin{itemize}
    \item \textbf{WavMark}\cite{chen2023wavmark}: WavMark is a robust and high-performance audio watermarking method based on Invertible Neural Networks.
    \item \textbf{AudioSeal}\cite{san2024proactive}: AudioSeal is a state-of-the-art (SOTA) audio watermarking method featuring an encoder-decoder symmetric architecture.
    \item \textbf{TraceableSpeech}\cite{zhou2024traceablespeech}: TraceableSpeech is an audio codec watermarking method that encodes watermarks into features and performs temporal broadcasting fusion with acoustic tokens post-quantization.
    \item \textbf{WMCodec}\cite{zhou2024wmcodec}: WMCodec is an SOTA audio codec watermarking method that encodes watermarks into features and performs temporal attention fusion with acoustic tokens pre-quantization.
\end{itemize}

\subsubsection{Evaluation Metric}

We use Bit-wise Accuracy (ACC) as a metric to assess decoding accuracy, which is defined within the range of [0, 1]. An ACC value of 0.5 indicates performance equivalent to random guessing. The calculation formula for ACC is as follows:
$$
\text{ACC}(w, \hat{w}) = \frac{1}{l} \sum_{i=1}^{l} \mathbb{I}(w_i = \hat{w}_i),
$$
where $l$ is the ground truth binary watermark, \( \hat{w} \) is the predicted binary watermark, and \( k \) is the length of the watermark sequence.

We employed Perceptual Evaluation of Speech Quality (PESQ)\cite{rix2001perceptual}, Short-Time Objective Intelligibility (STOI)\cite{taal2010short}, Mel distance (Mel Dis) and STFT distance (STFT Dis) as metrics to evaluate the quality of generated audio, which reflects the imperceptibility of the watermark in the audio after watermarking.

\subsubsection{Implementation Details}

For P2Mark-Vocoder, we use HiFi-GAN\footnote{https://github.com/jik876/hifi-gan} as the base model. We pre-trained HiFi-GAN using the training set of LibriTTS with the audio segment size set to 8192. During training, we used 8 V100 GPUs with a batch size of 128 and a learning rate of 0.0002, training for 300,000 iterations.

For P2Mark-Codec, we use HiFi-Codec as the base model. We pre-trained HiFi-Codec\footnote{https://github.com/yangdongchao/AcademiCodec} using the training set of LibriTTS, without grouped quantization, but with 8 layers of residual quantization, and the audio segment size set to 24000. During training, we used 8 V100 GPUs with a batch size of 128 and a learning rate of 0.0002, training for 400,000 iterations.

After obtaining the pre-trained HiFi-GAN and HiFi-Codec models, we replaced the 1-d convolution layers in the generators of HiFi-GAN and HiFi-Codec with 1-d convolution layers equipped with WM-LoRA and added a watermark encoder module and watermark extractors module for training. Subsequently, we used a single V100 GPU with a batch size of 16 for fine-tuning with watermark encoder and extraction modules.
% , training for an additional 100,000 iterations.

\subsection{Experimental Results}

In this section, we conducted a comprehensive evaluation of the proposed method P2Mark through extensive experiments. First, we compared our method with SOTA audio watermarking methods, including audio-level audio watermarking and feature-level audio watermarking. We compared the watermark extraction accuracy and the speech quality after the watermarking process of different methods. Then, we conducted ablation experiments to explore the impact of our proposed WGOPO optimization method and the watermark capacity of P2Mark. Finally, we tested the robustness of P2Mark under various attacks.

\begin{table}[t!]
	\centering
	\resizebox{\linewidth}{!}{
		\begin{tabular}{llcccccc|c}
			\hline
			\multirow{2}{*}{\textbf{Task}} & \multirow{2}{*}{\textbf{Method}} & \multirow{2}{*}{\textbf{Type}} & \multirow{2}{*}{\textbf{WB-P}} & \multicolumn{4}{c|}{\textbf{Audio quality metrics}} & \multirow{2}{*}{\textbf{ACC{\color{highlightred} $\uparrow$}}}\\ % \cline{2-8}
             
			&&&& PESQ{\color{highlightred} $\uparrow$} & STOI{\color{highlightred} $\uparrow$} & Mel Dis{\color{highlightgreen} $\downarrow$} & STFT Dis{\color{highlightgreen} $\downarrow$} &  \\ \hline
            
            % \rowcolor{shadecolor}
            \multirow{4}{*}{\textbf{Vocoder}} & HiFi-GAN  
            &&& 3.25 & 0.966 & 3.26 & 3.10 & -- \\ 
            \cdashline{2-9}
            
            &WavMark\cite{chen2023wavmark} 
            & Audio-level & \usym{2717}                 
            & 3.09                               & \textcolor{blue}{\textbf{0.964}} 
            & 3.94                               & 3.20 
            & \textcolor{deepred}{\textbf{1.00}} \\
            
            &AudioSeal\cite{san2024proactive} 
            & Audio-level & \usym{2717}            
            & \textcolor{blue}{\textbf{3.17}}    & \textcolor{deepred}{\textbf{0.965}} 
            & \textcolor{deepred}{\textbf{3.40}} & \textcolor{deepred}{\textbf{3.12}} 
            & \textcolor{deepred}{\textbf{1.00}} \\
            
            % \rowcolor{LightCyan1}
            \rowcolor{shadecolor}
            &P2Mark-Vocoder%(Ours) 
            & Parameter-level & \usym{2713}                           
            & \textcolor{deepred}{\textbf{3.21}} & \textcolor{deepred}{\textbf{0.965}} 
            & \textcolor{blue}{\textbf{3.46}}    & \textcolor{blue}{\textbf{3.19}} 
            & \textcolor{deepred}{\textbf{1.00}} \\
            
            \hline
            \hline
            
            % \rowcolor{shadecolor}
            \multirow{6}{*}{\textbf{Codec}} & HiFi-Codec
            &&& 3.52 & 0.966 & 3.02 & 2.71 & --   \\
            \cdashline{2-9}
            
            &WavMark\cite{chen2023wavmark} 
            & Audio-level & \usym{2717}                   
            & 3.32 & 0.963 
            & 3.69 & 2.82 & \textcolor{deepred}{\textbf{1.00}} \\
            
            &AudioSeal\cite{san2024proactive} 
            & Audio-level & \usym{2717}               
            & \textcolor{blue}{\textbf{3.45}}   & \textcolor{deepred}{\textbf{0.964}}
            & 3.20   & \textcolor{deepred}{\textbf{2.73}}
            & \textcolor{deepred}{\textbf{1.00}} \\
            
            &TraceableSpeech\cite{zhou2024traceablespeech} 
            % & \multirow{3}{*}{GM}
            & Feature-level & \usym{2717} 
            & 3.11 & 0.959 & 3.53 & 2.89 & \textcolor{deepred}{\textbf{1.00}} \\
            
            &WMCodec\cite{zhou2024wmcodec} 
            & Feature-level & \usym{2717} 
            & 3.43 & \textcolor{blue}{\textbf{0.961}} & \textcolor{blue}{\textbf{3.13}} & 2.77 & \textcolor{deepred}{\textbf{1.00}} \\
            
            % \rowcolor{LightCyan1}
            \rowcolor{shadecolor}
            &P2Mark-Codec%(Ours) 
            & Parameter-level & \usym{2713}                             
            & \textcolor{deepred}{\textbf{3.48}}& \textcolor{deepred}{\textbf{0.964}} 
            & \textcolor{deepred}{\textbf{3.09}}& \textcolor{blue}{\textbf{2.74}} 
            & \textcolor{deepred}{\textbf{1.00}} \\
            \hline
        \end{tabular}
    }
\caption{Performance comparison between two variants of P2Mark on speech generation models' decoders: P2Mark-Vocoder and P2Mark-Codec, against baseline audio watermarking models. 
% PH indicates Post-hoc audio watermarking method, and GM indicates generative model audio watermarking method. 
WB-P indicates whether the method can provide white box protection in the source code and weights open source scenario. The \textcolor{deepred}{\textbf{red}} denotes the highest result, and the \textcolor{blue}{\textbf{blue}} denotes the second highest result.}
\label{tab_main}
\end{table}

\subsubsection{Comparison Results with Baselines} 

Table \ref{tab_main} presents the comparison results of our method with other baseline methods. 
For audio watermarking applied to vocoders, we compared P2Mark-Vocoder with the audio-level audio watermarking methods WavMark\cite{chen2023wavmark} and AudioSeal\cite{san2024proactive}. The results indicate that all methods achieved an extraction accuracy of 1.00 for 16-bit binary watermarks. In terms of audio quality metrics, P2Mark-Vocoder outperformed the two post-hoc audio-level watermarking methods in PESQ and was only slightly inferior to AudioSeal in terms of Mel distance and STFT distance. 

For audio watermarking applied to codecs, we compared our P2Mark-Codec with two audio-level audio watermarking methods, WavMark and AudioSeal, as well as two feature-level audio watermarking methods, TraceableSpeech \cite{zhou2024traceablespeech} and WMCodec \cite{zhou2024wmcodec}. The results indicate that all methods achieved an extraction accuracy of 1.00 for 16-bit binary watermarks. In terms of audio quality metrics, P2Mark-Vocoder outperformed all four baseline methods in PESQ, STOI, and Mel distance, while being only slightly inferior to AudioSeal in STFT distance.

It is important to clarify that we do not claim that P2Mark achieves SOTA performance across all metrics, as previous methods could not offer flexible white-box protection with both codes and model weights being open source. Our method can be applied in scenarios suitable for baseline methods and achieves comparable performance, but baseline methods are not applicable in the white-box protection scenarios where our method can be employed.

\subsubsection{Ablation Study}
Our ablation studies systematically investigate two critical design factors: watermark capacity scaling and the efficacy of the proposed optimization method, WGOPO. 
% The results in Table \ref{ablation} reveal three key insights.

\textbf{Impact of Watermark Capacity Scaling on Performance.}
As shown in Table \ref{ablation}, increasing the watermark payload from 16 bits to 32 bits, the watermark extraction accuracy for both P2Mark-Vocoder and P2Mark-Codec remains at 1.00. This demonstrates that P2Mark can be further scaled to higher watermark capacities. However, as the watermark capacity increases, the quality of the generated speech gradually decreases. For P2Mark-Vocoder, this expansion results in a decrease of 0.17 in PESQ and an increase of 0.34 in Mel distance. P2Mark-Codec shows a similar trend, with a decrease of 0.06 in PESQ and an increase of 0.05 in Mel distance. Overall, as the number of embedded watermark bits increases, the quality of the audio generated by P2Mark declines slightly, but the embedding and extraction of the watermark remain effective, indicating the scalability of our method.

\begin{table}[t!]
    \centering
    \resizebox{\linewidth}{!}{
        \begin{tabular}{llccccc|c}
        \hline
        \multirow{2}{*}{\textbf{Task}} &\multirow{2}{*}{\textbf{Variant}} & \multirow{2}{*}{\textbf{Bits}} & \multicolumn{4}{c|}{\textbf{Audio quality metrics}} & \multirow{2}{*}{\textbf{ACC{\color{highlightred} $\uparrow$}}} \\ % \cline{2-8}
        & & & PESQ{\color{highlightred} $\uparrow$} & STOI{\color{highlightred} $\uparrow$} & Mel Dis{\color{highlightgreen} $\downarrow$} & STFT Dis{\color{highlightgreen} $\downarrow$} & \\ \hline
        % \rowcolor{shadecolor}
        \multirow{5}{*}{\textbf{Vocoder}}&HiFi-GAN                & 
        & 3.25          & 0.966         & 3.26          & 3.10          & -- \\ 
        \cdashline{2-8}
        \rowcolor{shadecolor}
        &P2Mark-Vocoder          & \multirow{2}{*}{16} 
        & 3.21          & 0.965         & 3.46          & 3.19          & 1.00 \\
        &\ \ \ -\ w/o\ WGOPO& 
        & 3.18(-0.03)   & 0.959(-0.006) & 3.60(+0.14)   & 3.22(+0.03)   & 1.00(-0.00) \\
        \cdashline{2-8}
        &P2Mark-Vocoder          & \multirow{2}{*}{32} 
        & 3.04          & 0.955         & 3.80          & 3.29          & 1.00 \\
        &\ \ \ -\ w/o\ WGOPO& 
        & 2.94(-0.10)   & 0.947(-0.008) & 3.98(+0.18)   & 3.32(+0.03)   & 0.97(-0.03) \\
        \hline
        \hline
        % \rowcolor{shadecolor}
        \multirow{5}{*}{\textbf{Codec}} &   HiFi-Codec  & 
        & 3.52          & 0.966         & 3.02          & 2.71          & -- \\
        \cdashline{2-8}
        \rowcolor{shadecolor}
        &P2Mark-Codec            & \multirow{2}{*}{16} 
        & 3.48          & 0.964         & 3.09          & 2.74          & 1.00 \\
        &\ \ \ -\ w/o\ WGOPO& 
        & 3.36(-0.12)   & 0.960(-0.004) & 3.21(+0.12)   & 2.78(+0.04)   & 0.98(-0.02) \\
        \cdashline{2-8}
        &P2Mark-Codec            & \multirow{2}{*}{32} 
        & 3.42          & 0.963         & 3.14          & 2.75          & 1.00  \\
        &\ \ \ -\ w/o\ WGOPO& 
        & 3.29(-0.13)   & 0.957(-0.006) & 3.33(+0.19)   & 2.81(+0.06)   & 0.99(-0.01)  \\
        \hline
        \end{tabular}}
    \caption{The ablation study on the efficiency of WGOPO and the watermark capacity.}
    \label{ablation}
\end{table}

\textbf{The Effectiveness of WGOPO.}
As a gradient optimization method for P2Mark, WGOPO demonstrates significant effectiveness in enhancing performance. As shown in Table \ref{ablation}, the removal of WGOPO leads to a consistent decline in performance across various configurations. Specifically, for the 16-bit P2Mark-Vocoder, the absence of WGOPO results in a decrease of 0.03 in the PESQ and an increase of 0.14 in the Mel distance. Similarly, for P2Mark-Codec, the absence of WGOPO results in a decrease of 0.12 in the PESQ and an increase of 0.12 in the Mel distance for 16-bit watermarking scenarios. This result confirms our hypothesis that WGOPO, by effectively decoupling watermarking from speech quality optimization, not only ensures better watermark fusion but also more effectively optimizes the generator. This approach minimizes the degradation in generative performance caused by watermark injection in parameters. 

Notably, WGOPO gains importance as the watermark complexity increases. Without WGOPO, the performance degradation in various audio quality evaluation metrics is greater for a 32-bit watermark compared to a 16-bit watermark. This indicates that higher capacity watermarks require more sophisticated optimization to maintain their stealthiness. The decrease of watermarking detection ACC at 32 bits further indicates that WGOPO helps maintain watermark integrity under capacity pressure.
These findings collectively validate our core design philosophy: parameter fusion requires co-optimization mechanisms like WGOPO to achieve secure yet imperceptible watermarking.

\begin{table}[t!]
    \centering
    \resizebox{\linewidth}{!}{
    \begin{tabular}{llp{9cm}}
    \toprule
    \textbf{Attack Type} & \textbf{Subtype} & \textbf{Description} \\
    \midrule
    \multirow{2}{*}{Noise} 
        & Pink & Adds pink noise to audio signal (std=0.1) \\
        & White & Adds Gaussian noise to audio signal (std=0.05) \\
    \midrule
    \multirow{3}{*}{Filtering}
        & Lowpass  & Applies lowpass filter with 500 Hz cutoff \\
        & Bandpass  & Applies Bandpass filtering in 500 Hz - 1.5 kHz \\
        & Highpass  & Applies highpass filter with 1.5 kHz cutoff \\
    \midrule
    \multirow{2}{*}{Volume}
        & Boost & Amplifies audio by factor 10 \\
        & Duck  & Reduces volume by factor 0.1 \\
    \midrule
    \multirow{2}{*}{Compression}
        & MP3 & MP3 codec at 128 kbps bitrate \\
        & AAC & AAC codec at 128 kbps bitrate \\
    \midrule
    \multirow{3}{*}{Others}
        & Resample & Upsamples from 24 kHz to 44.1 kHz then downsamples back \\
        & Echo & Adds 0.5s delay with 0.5 decay factor \\
        & Crop & Keeps only the first half of waveform \\
    \bottomrule
    \end{tabular}
    }
    \caption{Detailed description of audio attack types and their settings.}
    \label{tab:attack_descriptions}
\end{table}

\subsubsection{Robustness Evaluation Results}

To evaluate the robustness of the watermark, we subjected the generated audio to various robustness attacks: noise (pink noise, white noise), filtering (lowpass, bandpass, highpass), audio volume (boost audio, duck audio), compression (MP3, AAC), and other editing operations (resample, echo, crop). The details of the attacks are as Table \ref{tab:attack_descriptions}.

\begin{table}[t!]
    \centering
    \resizebox{\linewidth}{!}{
    \begin{tabular}{llcccc}
    \hline
    \multirow{2}{*}{\textbf{Attack Type}} & \multirow{2}{*}{\textbf{Subtype}} & \multicolumn{4}{c}{\textbf{Method}} \\
    \cmidrule(lr){3-6}
    & & \textbf{WavMark} & \textbf{AudioSeal} & \textbf{P2Mark-Vocoder} & \textbf{P2Mark-Codec} \\
    \hline
    None         &           & 1.00 & 1.00 & 1.00 & 1.00 \\
    \cdashline{1-6}
    \multirow{2}{*}{Noise} 
                 & Pink     & 0.98 & 0.99 & 0.98 & 0.99 \\
                 & White    & \underline{0.50} & \underline{0.62} & \underline{0.60} & \underline{0.55} \\
    \cdashline{1-6}
    \multirow{3}{*}{Filtering}
                 & Lowpass      & \underline{0.50} & \underline{0.50} & \underline{0.50} & \underline{0.50} \\
                 & Bandpass     & \underline{0.50} & 1.00 & \underline{0.76} & \underline{0.72} \\
                 & Highpass     & 1.00 & \underline{0.49} & 0.99 & 1.00 \\
    \cdashline{1-6}
    \multirow{2}{*}{Volume}
                 & Boost     & 1.00 & 1.00 & 1.00 & 1.00 \\
                 & Duck      & 1.00 & 1.00 & 1.00 & 1.00 \\
    \cdashline{1-6}
    \multirow{2}{*}{Compression}
                 & MP3       & 1.00 & 1.00 & 0.98 & 0.99 \\
                 & AAC       & 1.00 & \underline{0.63} & 1.00 & 1.00 \\
    \cdashline{1-6}
    \multirow{3}{*}{Others}
                 & Resample  & 1.00 & 1.00 & 1.00 & 1.00 \\
                 & Echo      & 0.97 & 1.00 & 1.00 & 1.00 \\
                 & Crop      & 0.96 & 1.00 & 1.00 & 1.00 \\
    \hline
    \end{tabular}
    }
    \caption{Robustness comparison under various attacks. The \underline{underline} indicates a watermark extraction accuracy below 0.90.}
    \label{tab:robust_results}
\end{table}

Existing audio watermarking methods typically enhance the robustness of watermarks against various attacks by incorporating simulated attacks during training. However, this simulation approach during training struggles to cover all types of attacks that may occur in real-world scenarios. Unlike previous methods, our approach integrates watermark embedding at a parameter level, inherently providing a certain degree of robustness. Remarkably, P2Mark still demonstrates excellent robustness without any simulated attacks during training.
Table \ref{tab:robust_results} shows the watermark extraction results of our method compared to baseline methods WavMark and AudioSeal when facing multiple attacks. All three methods perform poorly against white noise and low-pass attacks. Besides, WavMark and our method are sensitive to band-pass attacks, while AudioSeal is sensitive to high-pass and AAC attacks. Overall, P2Mark exhibits robustness comparable to SOTA watermarking methods.

\subsection{Discussion}
Our proposed P2Mark offers a reliable solution for watermark protection of NSG models in open-source settings, where both the model and its source code are publicly available. This scenario primarily targets proactive tracing of speech generated by open-source models and copyright protection of the models themselves. P2Mark simultaneously achieves both security and flexibility for parameter-level watermarking.
Security refers to the fact that, after the model and source code are open-sourced, P2Mark is significantly more resistant to circumvention, tampering, or removal compared to traditional audio-level and feature-level watermarking methods. Flexibility refers to the ability to freely modify the embedded watermark content without retraining the model before its release, unlike two concurrent parameter-level watermarking methods. This capability prevents watermark leakage and supports effective version control.

We comprehensively compare P2Mark with several SOTA open-source audio watermarking methods in terms of watermark imperceptibility, extraction accuracy, and robustness. First, P2Mark achieves evaluation results on par with SOTA audio-level and feature-level watermarking methods in terms of watermark imperceptibility, particularly under PESQ and STOI metrics. This is largely attributed to our plug-in fine-tuning on pretrained NSG models and the proposed WGOPO optimization strategy.
Second, in terms of watermark extraction accuracy, our experiments embedding random 16-bit binary watermarks into speech samples ranging from 1 to 10 seconds demonstrate that P2Mark achieves a 1.00 extraction accuracy, comparable to SOTA audio-level and feature-level methods. Given that the main purpose of protecting open-source models is proactive tracing and copyright verification, watermark capacity is not the primary focus; a 16-bit watermark is already sufficient for reliable identification.
Third, regarding robustness, we surprisingly observe that even without explicitly simulating preset attacks during training, P2Mark achieves robustness comparable to previous methods. We attribute this to the deep integration of the watermark into the model parameters. However, low-pass filtering causes catastrophic degradation of watermark detection, suggesting that the embedded watermark information likely resides in the high-frequency components of the generated audio.
Our ablation studies further show that increasing the watermark capacity to 32 bits leads to a slight degradation in audio quality but still maintains accurate extraction. Moreover, the benefits of the WGOPO optimization method become even more pronounced at higher watermark capacities.

In summary, P2Mark enables flexible open-source white-box protection for NSG, a capability previously unattainable with existing methods, while maintaining competitive performance in watermark imperceptibility, extraction accuracy, and robustness. These results highlight the advantages of our proposed approach.

\section{Conclusion}

This paper addresses the critical challenge of protecting open-source NSG systems where codes and model weights are fully public. 
We propose P2Mark, a plug-and-play parameter-level watermarking method for NSG. 
First, our plug-and-play watermarking module enables easy integration with mainstream waveform decoders, vocoder and codec, for NSG. 
Secondly, the parameter fusion mechanism permanently embeds watermarks into the model weights by the watermark adapter merging process. This process allows for flexible modification of the watermark content during the adapter merging. Once merged, the watermark cannot be removed or manipulated by simply altering the code.
Third, the watermarking gradient orthogonal projection optimization reduces the mutual interference between the watermarking and the generating optimization, ensuring the imperceptibility of the watermark and the accuracy of the watermark extraction.
Experiments have demonstrated the performance of P2Mark comparable to existing SOTA audio-level watermarking methods and feature-level watermarking methods in audio quality, watermark extraction accuracy, and robustness.
Crucially, P2Mark enables proactive tracing and copyright protection when both the NSG model and its source code are open-sourced, making it difficult for users to bypass the watermark embedding process or remove or manipulate the embedded watermark information through code modifications.
% This approach is of significant importance for the prevention of security risks and the protection of copyrights for open-source NSG models.

\section{Limitations and Future Work}

Although P2Mark demonstrates effective plug-and-play open-source protection for NSG, along with promising performance in terms of watermark extraction accuracy and the quality of generated audio, there are several limitations that need to be addressed.
First, as the watermark capacity increases, the complexity of training also grows, and the model struggles to converge when attempting to extend the watermark capacity to 64 bits. Overcoming this challenge and finding ways to further enhance the watermarking capacity is a critical avenue for future research.
Second, while P2Mark shows robustness against various attacks, evaluating its defense capability against white-box adversarial attacks remains an essential challenge. Developing effective defense mechanisms against potential adversarial threats is a key focus for future work.
Moving forward, we aim to explore more advanced techniques for parameter fusion and optimization, as well as investigate novel defense strategies against white-box adversarial attacks. These efforts will not only enhance the watermark capacity but also improve the robustness and security of watermarking techniques in NSG.

% Despite P2Mark's ability to implement flexible plug-and-play white-box protection for  and its good performance in watermark extraction accuracy and generated audio quality, there are still some limitations. First, as the watermark capacity increases, the difficulty of training also increases, and it becomes impossible to converge when expanding further to a 64-bit watermark. How to further increase the capacity of the watermark remains a future research goal. Secondly, evaluating the defense capability of P2Mark against white-box adversarial attacks and developing effective defenses against potential adversarial threats is also a key research goal for the future. In the future, we will further explore more effective parameter fusion and optimization methods, as well as defense strategies against white-box adversarial attacks, to enhance the capacity and robustness of watermarking techniques.
% Secondly, although P2Mark allows for changing different watermark contents after the watermark Adapter is trained, integrating them into the model parameters without retraining, the number of watermark bits is fixed. The method of embedding watermarks with variable bit numbers is also a future research target.

\section*{Acknowledgements}
This work is supported by the Scientific and Technological Innovation Important Plan of China (No. 2021ZD0201502), the National Natural Science Foundation of China (NSFC) (No. 62322120, No.U21B2010, No. 62306316, No. 62206278).

\bibliographystyle{elsarticle-num} 
\bibliography{ref.bib}

\begin{thebibliography}{10}
\expandafter\ifx\csname url\endcsname\relax
  \def\url#1{\texttt{#1}}\fi
\expandafter\ifx\csname urlprefix\endcsname\relax\def\urlprefix{URL }\fi
\expandafter\ifx\csname href\endcsname\relax
  \def\href#1#2{#2} \def\path#1{#1}\fi

\bibitem{du2024cosyvoice}
Z.~Du, Q.~Chen, S.~Zhang, K.~Hu, H.~Lu, Y.~Yang, H.~Hu, S.~Zheng, Y.~Gu, Z.~Ma, et~al., Cosyvoice: A scalable multilingual zero-shot text-to-speech synthesizer based on supervised semantic tokens, arXiv preprint arXiv:2407.05407 (2024).

\bibitem{du2024cosyvoice2}
Z.~Du, Y.~Wang, Q.~Chen, X.~Shi, X.~Lv, T.~Zhao, Z.~Gao, Y.~Yang, C.~Gao, H.~Wang, et~al., Cosyvoice 2: Scalable streaming speech synthesis with large language models, arXiv preprint arXiv:2412.10117 (2024).

\bibitem{wang2024maskgct}
Y.~Wang, H.~Zhan, L.~Liu, R.~Zeng, H.~Guo, J.~Zheng, Q.~Zhang, X.~Zhang, S.~Zhang, Z.~Wu, Maskgct: Zero-shot text-to-speech with masked generative codec transformer, arXiv preprint arXiv:2409.00750 (2024).

\bibitem{wang2025sparktts}
X.~Wang, M.~Jiang, Z.~Ma, Z.~Zhang, S.~Liu, L.~Li, Z.~Liang, Q.~Zheng, R.~Wang, X.~Feng, W.~Bian, Z.~Ye, S.~Cheng, R.~Yuan, Z.~Zhao, X.~Zhu, J.~Pan, L.~Xue, P.~Zhu, Y.~Chen, Z.~Li, X.~Chen, L.~Xie, Y.~Guo, W.~Xue, \href{https://arxiv.org/abs/2503.01710}{Spark-tts: An efficient llm-based text-to-speech model with single-stream decoupled speech tokens} (2025).
\newblock \href {http://arxiv.org/abs/2503.01710} {\path{arXiv:2503.01710}}.
\newline\urlprefix\url{https://arxiv.org/abs/2503.01710}

\bibitem{yamagishi2021asvspoof}
J.~Yamagishi, X.~Wang, M.~Todisco, M.~Sahidullah, J.~Patino, A.~Nautsch, X.~Liu, K.~A. Lee, T.~Kinnunen, N.~Evans, et~al., Asvspoof 2021: accelerating progress in spoofed and deepfake speech detection, 2021 Edition of the Automatic Speaker Verification and Spoofing Countermeasures Challenge (2021).

\bibitem{yi2022add}
J.~Yi, R.~Fu, J.~Tao, S.~Nie, H.~Ma, C.~Wang, T.~Wang, Z.~Tian, Y.~Bai, C.~Fan, et~al., Add 2022: the first audio deep synthesis detection challenge, in: ICASSP 2022-2022 IEEE International Conference on Acoustics, Speech and Signal Processing (ICASSP), IEEE, 2022, pp. 9216--9220.

\bibitem{yi2023add}
J.~Yi, J.~Tao, R.~Fu, X.~Yan, C.~Wang, T.~Wang, C.~Y. Zhang, X.~Zhang, Y.~Zhao, Y.~Ren, et~al., Add 2023: the second audio deepfake detection challenge, in: CEUR Workshop Proceedings, Vol. 3597, 2023, pp. 125--130.

\bibitem{o2024maskmark}
P.~O’Reilly, Z.~Jin, J.~Su, B.~Pardo, Maskmark: Robust neuralwatermarking for real and synthetic speech, in: ICASSP 2024-2024 IEEE International Conference on Acoustics, Speech and Signal Processing (ICASSP), IEEE, 2024, pp. 4650--4654.

\bibitem{liu2023detecting}
C.~Liu, J.~Zhang, T.~Zhang, X.~Yang, W.~Zhang, N.~Yu, Detecting voice cloning attacks via timbre watermarking, arXiv preprint arXiv:2312.03410 (2023).

\bibitem{liu2023dear}
C.~Liu, J.~Zhang, H.~Fang, Z.~Ma, W.~Zhang, N.~Yu, Dear: A deep-learning-based audio re-recording resilient watermarking, in: Proceedings of the AAAI Conference on Artificial Intelligence, Vol.~37, 2023, pp. 13201--13209.

\bibitem{chen2023wavmark}
G.~Chen, Y.~Wu, S.~Liu, T.~Liu, X.~Du, F.~Wei, Wavmark: Watermarking for audio generation, arXiv preprint arXiv:2308.12770 (2023).

\bibitem{san2024proactive}
R.~San~Roman, P.~Fernandez, H.~Elsahar, A.~D{\'e}fossez, T.~Furon, T.~Tran, Proactive detection of voice cloning with localized watermarking, in: International Conference on Machine Learning, Vol. 235, 2024.

\bibitem{liu2024groot}
W.~Liu, Y.~Li, D.~Lin, H.~Tian, H.~Li, Groot: Generating robust watermark for diffusion-model-based audio synthesis, arXiv preprint arXiv:2407.10471 (2024).

\bibitem{zhou2024traceablespeech}
J.~Zhou, J.~Yi, T.~Wang, J.~Tao, Y.~Bai, C.~Y. Zhang, Y.~Ren, Z.~Wen, Traceablespeech: Towards proactively traceable text-to-speech with watermarking, arXiv preprint arXiv:2406.04840 (2024).

\bibitem{zhou2024wmcodec}
J.~Zhou, J.~Yi, Y.~Ren, J.~Tao, T.~Wang, C.~Y. Zhang, Wmcodec: End-to-end neural speech codec with deep watermarking for authenticity verification, arXiv preprint arXiv:2409.12121 (2024).

\bibitem{wang2021k}
R.~Wang, D.~Tang, N.~Duan, Z.~Wei, X.-J. Huang, J.~Ji, G.~Cao, D.~Jiang, M.~Zhou, K-adapter: Infusing knowledge into pre-trained models with adapters, in: Findings of the Association for Computational Linguistics: ACL-IJCNLP 2021, 2021, pp. 1405--1418.

\bibitem{hu2021lora}
E.~J. Hu, Y.~Shen, P.~Wallis, Z.~Allen-Zhu, Y.~Li, S.~Wang, L.~Wang, W.~Chen, Lora: Low-rank adaptation of large language models, arXiv preprint arXiv:2106.09685 (2021).

\bibitem{van2016wavenet}
S.~Dieleman, H.~Zen, K.~Simonyan, O.~Vinyals, A.~Graves, N.~Kalchbrenner, A.~Senior, K.~Kavukcuoglu, et~al., Wavenet: A generative model for raw audio, arXiv preprint arXiv:1609.03499 12 (2016).

\bibitem{wang2017tacotron}
Y.~Wang, R.~Skerry-Ryan, D.~Stanton, Y.~Wu, R.~J. Weiss, N.~Jaitly, Z.~Yang, Y.~Xiao, Z.~Chen, S.~Bengio, et~al., Tacotron: Towards end-to-end speech synthesis, Interspeech 2017 (2017) 4006.

\bibitem{shen2018natural}
J.~Shen, R.~Pang, R.~J. Weiss, M.~Schuster, N.~Jaitly, Z.~Yang, Z.~Chen, Y.~Zhang, Y.~Wang, R.~Skerrv-Ryan, et~al., Natural tts synthesis by conditioning wavenet on mel spectrogram predictions, in: 2018 IEEE international conference on acoustics, speech and signal processing (ICASSP), IEEE, 2018, pp. 4779--4783.

\bibitem{ren2019fastspeech}
Y.~Ren, Y.~Ruan, X.~Tan, T.~Qin, S.~Zhao, Z.~Zhao, T.-Y. Liu, Fastspeech: Fast, robust and controllable text to speech, Advances in neural information processing systems 32 (2019).

\bibitem{renfastspeech}
Y.~Ren, C.~Hu, X.~Tan, T.~Qin, S.~Zhao, Z.~Zhao, T.-Y. Liu, Fastspeech 2: Fast and high-quality end-to-end text to speech, in: International Conference on Learning Representations, 2021.

\bibitem{popov2021grad}
V.~Popov, I.~Vovk, V.~Gogoryan, T.~Sadekova, M.~Kudinov, Grad-tts: A diffusion probabilistic model for text-to-speech, in: International Conference on Machine Learning, PMLR, 2021, pp. 8599--8608.

\bibitem{kim2020glow}
J.~Kim, S.~Kim, J.~Kong, S.~Yoon, Glow-tts: A generative flow for text-to-speech via monotonic alignment search, Advances in Neural Information Processing Systems 33 (2020) 8067--8077.

\bibitem{kim2021conditional}
J.~Kim, J.~Kong, J.~Son, Conditional variational autoencoder with adversarial learning for end-to-end text-to-speech, in: International Conference on Machine Learning, PMLR, 2021, pp. 5530--5540.

\bibitem{wang2023neural}
C.~Wang, S.~Chen, Y.~Wu, Z.~Zhang, L.~Zhou, S.~Liu, Z.~Chen, Y.~Liu, H.~Wang, J.~Li, et~al., Neural codec language models are zero-shot text to speech synthesizers, arXiv preprint arXiv:2301.02111 (2023).

\bibitem{lajszczak2024base}
M.~{\L}ajszczak, G.~C{\'a}mbara, Y.~Li, F.~Beyhan, A.~van Korlaar, F.~Yang, A.~Joly, {\'A}.~Mart{\'\i}n-Cortinas, A.~Abbas, A.~Michalski, et~al., Base tts: Lessons from building a billion-parameter text-to-speech model on 100k hours of data, arXiv preprint arXiv:2402.08093 (2024).

\bibitem{anastassiou2024seed}
P.~Anastassiou, J.~Chen, J.~Chen, Y.~Chen, Z.~Chen, Z.~Chen, J.~Cong, L.~Deng, C.~Ding, L.~Gao, et~al., Seed-tts: A family of high-quality versatile speech generation models, arXiv preprint arXiv:2406.02430 (2024).

\bibitem{kim2024clam}
J.~Kim, K.~Lee, S.~Chung, J.~Cho, Clam-tts: Improving neural codec language model for zero-shot text-to-speech, arXiv preprint arXiv:2404.02781 (2024).

\bibitem{huang24_interspeech}
F.~Huang, K.~Zeng, W.~Zhu, Diffvc+: Improving diffusion-based voice conversion for speaker anonymization, in: Interspeech 2024, 2024, pp. 4453--4457.
\newblock \href {https://doi.org/10.21437/Interspeech.2024-502} {\path{doi:10.21437/Interspeech.2024-502}}.

\bibitem{baade2024neural}
A.~Baade, P.~Peng, D.~Harwath, Neural codec language models for disentangled and textless voice conversion, in: Proc. Interspeech 2024, 2024, pp. 182--186.

\bibitem{wang2024emotion}
T.~Wang, J.~Yi, R.~Fu, J.~Tao, Z.~Wen, C.~Y. Zhang, Emotion selectable end-to-end text-based speech editing, Artificial Intelligence 329 (2024) 104076.

\bibitem{peng2024voicecraft}
P.~Peng, P.-Y. Huang, S.-W. Li, A.~Mohamed, D.~Harwath, Voicecraft: Zero-shot speech editing and text-to-speech in the wild, arXiv preprint arXiv:2403.16973 (2024).

\bibitem{prenger2019waveglow}
R.~Prenger, R.~Valle, B.~Catanzaro, Waveglow: A flow-based generative network for speech synthesis, in: ICASSP 2019-2019 IEEE International Conference on Acoustics, Speech and Signal Processing (ICASSP), IEEE, 2019, pp. 3617--3621.

\bibitem{kumar2019melgan}
K.~Kumar, R.~Kumar, T.~De~Boissiere, L.~Gestin, W.~Z. Teoh, J.~Sotelo, A.~De~Brebisson, Y.~Bengio, A.~C. Courville, Melgan: Generative adversarial networks for conditional waveform synthesis, Advances in neural information processing systems 32 (2019).

\bibitem{kong2020hifi}
J.~Kong, J.~Kim, J.~Bae, Hifi-gan: Generative adversarial networks for efficient and high fidelity speech synthesis, Advances in neural information processing systems 33 (2020) 17022--17033.

\bibitem{lee2022bigvgan}
S.-g. Lee, W.~Ping, B.~Ginsburg, B.~Catanzaro, S.~Yoon, Bigvgan: A universal neural vocoder with large-scale training, arXiv preprint arXiv:2206.04658 (2022).

\bibitem{song2020bridging}
J.~Song, S.~Ermon, Bridging the gap between f-gans and wasserstein gans, in: International Conference on Machine Learning, Pmlr, 2020, pp. 9078--9087.

\bibitem{zeghidour2022soundstream}
N.~Zeghidour, A.~Luebs, A.~Omran, J.~Skoglund, M.~Tagliasacchi, Soundstream: An end-to-end neural audio codec, IEEE/ACM Transactions on Audio, Speech, and Language Processing 30 (2022) 495--507.

\bibitem{defossezhigh}
A.~D{\'e}fossez, J.~Copet, G.~Synnaeve, Y.~Adi, High fidelity neural audio compression, Transactions on Machine Learning Research (2023).

\bibitem{yang2023hifi}
D.~Yang, S.~Liu, R.~Huang, J.~Tian, C.~Weng, Y.~Zou, Hifi-codec: Group-residual vector quantization for high fidelity audio codec, arXiv preprint arXiv:2305.02765 (2023).

\bibitem{ren2024fewer}
Y.~Ren, T.~Wang, J.~Yi, L.~Xu, J.~Tao, C.~Y. Zhang, J.~Zhou, Fewer-token neural speech codec with time-invariant codes, in: ICASSP 2024-2024 IEEE International Conference on Acoustics, Speech and Signal Processing (ICASSP), IEEE, 2024, pp. 12737--12741.

\bibitem{kumar2024high}
R.~Kumar, P.~Seetharaman, A.~Luebs, I.~Kumar, K.~Kumar, High-fidelity audio compression with improved rvqgan, Advances in Neural Information Processing Systems 36 (2024).

\bibitem{zhang2024speechtokenizer}
X.~Zhang, D.~Zhang, S.~Li, Y.~Zhou, X.~Qiu, Speechtokenizer: Unified speech tokenizer for speech language models, in: ICLR, 2024.

\bibitem{gao2010geometric}
X.~Gao, C.~Deng, X.~Li, D.~Tao, Geometric distortion insensitive image watermarking in affine covariant regions, IEEE Transactions on Systems, Man, and Cybernetics, Part C (Applications and Reviews) 40~(3) (2010) 278--286.

\bibitem{patel2011unified}
S.~B. Patel, T.~B. Mehta, S.~N. Pradhan, A unified technique for robust digital watermarking of colour images using data mining and dct, International Journal of Internet Technology and Secured Transactions 3~(1) (2011) 81--96.

\bibitem{agarwal2019survey}
N.~Agarwal, A.~K. Singh, P.~K. Singh, Survey of robust and imperceptible watermarking, Multimedia Tools and Applications 78 (2019) 8603--8633.

\bibitem{rahman2013dwt}
M.~M. Rahman, A dwt, dct and svd based watermarking technique to protect the image piracy, International Journal of Managing Public Sector Information and Communication Technologies 4~(2) (2013) 21.

\bibitem{zhang2019robust}
K.~A. Zhang, L.~Xu, A.~Cuesta-Infante, K.~Veeramachaneni, Robust invisible video watermarking with attention, arXiv preprint arXiv:1909.01285 (2019).

\bibitem{zhou2022robust}
Y.~Zhou, Q.~Ying, Y.~Wang, X.~Zhang, Z.~Qian, X.~Zhang, Robust watermarking for video forgery detection with improved imperceptibility and robustness, in: 2022 IEEE 24th International Workshop on Multimedia Signal Processing (MMSP), IEEE, 2022, pp. 1--6.

\bibitem{ren2024copyright}
J.~Ren, H.~Xu, P.~He, Y.~Cui, S.~Zeng, J.~Zhang, H.~Wen, J.~Ding, P.~Huang, L.~Lyu, et~al., Copyright protection in generative ai: A technical perspective, arXiv preprint arXiv:2402.02333 (2024).

\bibitem{juvela2025audio}
L.~Juvela, X.~Wang, Audio codec augmentation for robust collaborative watermarking of speech synthesis, in: ICASSP 2025-2025 IEEE International Conference on Acoustics, Speech and Signal Processing (ICASSP), IEEE, 2025, pp. 1--5.

\bibitem{feng2024aqualora}
W.~Feng, W.~Zhou, J.~He, J.~Zhang, T.~Wei, G.~Li, T.~Zhang, W.~Zhang, N.~Yu, Aqualora: Toward white-box protection for customized stable diffusion models via watermark lora, arXiv preprint arXiv:2405.11135 (2024).

\bibitem{boney1996digital}
L.~Boney, A.~H. Tewfik, K.~N. Hamdy, Digital watermarks for audio signals, in: Proceedings of the third IEEE international conference on multimedia computing and systems, IEEE, 1996, pp. 473--480.

\bibitem{zhang2020time}
H.~Zhang, A time-frequency perspective on audio watermarking, arXiv preprint arXiv:2002.03156 (2020).

\bibitem{hu2020selection}
Y.~Hu, M.~Ma, W.~Lu, N.~N. Xiong, J.~Wei, Selection of the optimal embedding positions of digital audio watermarking in wavelet domain, arXiv preprint arXiv:2010.11461 (2020).

\bibitem{qin2023lattice}
J.~Qin, S.~Lyu, J.~Deng, X.~Liang, S.~Xiang, H.~Chen, A lattice-based embedding method for reversible audio watermarking, IEEE Transactions on Dependable and Secure Computing 21~(4) (2023) 2619--2630.

\bibitem{san2024latent}
R.~San~Roman, P.~Fernandez, A.~Deleforge, Y.~Adi, R.~Serizel, Latent watermarking of audio generative models, arXiv e-prints (2024) arXiv--2409.

\bibitem{cheng2024hifi}
X.~Cheng, Y.~Wang, C.~Liu, D.~Hu, Z.~Su, Hifi-ganw: Watermarked speech synthesis via fine-tuning of hifi-gan, IEEE Signal Processing Letters (2024).

\bibitem{casanova2022yourtts}
E.~Casanova, J.~Weber, C.~D. Shulby, A.~C. Junior, E.~G{\"o}lge, M.~A. Ponti, Yourtts: Towards zero-shot multi-speaker tts and zero-shot voice conversion for everyone, in: International Conference on Machine Learning, PMLR, 2022, pp. 2709--2720.

\bibitem{he2016deep}
K.~He, X.~Zhang, S.~Ren, J.~Sun, Deep residual learning for image recognition, in: Proceedings of the IEEE conference on computer vision and pattern recognition, 2016, pp. 770--778.

\bibitem{wu2024codec}
H.~Wu, H.-L. Chung, Y.-C. Lin, Y.-K. Wu, X.~Chen, Y.-C. Pai, H.-H. Wang, K.-W. Chang, A.~Liu, H.-Y. Lee, Codec-superb: An in-depth analysis of sound codec models, in: Findings of the Association for Computational Linguistics ACL 2024, 2024, pp. 10330--10348.

\bibitem{chaudhryefficient}
A.~Chaudhry, M.~Ranzato, M.~Rohrbach, M.~Elhoseiny, Efficient lifelong learning with a-gem, in: International Conference on Learning Representations, 2018.

\bibitem{zen2019libritts}
H.~Zen, V.~Dang, R.~Clark, Y.~Zhang, R.~J. Weiss, Y.~Jia, Z.~Chen, Y.~Wu, Libritts: A corpus derived from librispeech for text-to-speech, Interspeech 2019 (2019).

\bibitem{rix2001perceptual}
A.~W. Rix, J.~G. Beerends, M.~P. Hollier, A.~P. Hekstra, Perceptual evaluation of speech quality (pesq)-a new method for speech quality assessment of telephone networks and codecs, in: 2001 IEEE international conference on acoustics, speech, and signal processing. Proceedings (Cat. No. 01CH37221), Vol.~2, IEEE, 2001, pp. 749--752.

\bibitem{taal2010short}
C.~H. Taal, R.~C. Hendriks, R.~Heusdens, J.~Jensen, A short-time objective intelligibility measure for time-frequency weighted noisy speech, in: 2010 IEEE international conference on acoustics, speech and signal processing, IEEE, 2010, pp. 4214--4217.

\end{thebibliography}

\end{document}